\documentclass[twocolumn,nolinenumbers]{aastex701}
\usepackage{amsmath}
\usepackage{amssymb}
\usepackage{amsthm}
\usepackage{natbib,aasdefs,url,bm}
\usepackage{array}
\usepackage{float}
\usepackage{graphicx}
\usepackage{subfigure}
\usepackage[dvipsnames]{xcolor}
\hypersetup{linkcolor=red,citecolor=NavyBlue,filecolor=cyan,urlcolor=purple}

\def\be{\begin{equation}}
\def\ee{\end{equation}}
\def\ba{\begin{eqnarray}}
\def\ea{\end{eqnarray}}

\def\aap{\ {A\&A}\ }

\def\apj{\ {ApJ}\ }
\def\apjl{\ {ApJL}\ }
\def\apjs{\ {ApJS}\ }

\submitjournal{\apj}
\shorttitle{Intrinsic X-Ray-Selected Hidden Neutrino Active Galaxy Candidates (X-HiNAC)}
\shortauthors{Mukhopadhyay et al.}
\begin{document}
\title{Hidden Neutrino Active Galaxy Candidates: Prospects for Catalog and Stacking Searches with IceCube, IceCube-Gen2, and KM3NeT}
\correspondingauthor{Mainak Mukhopadhyay, Jose Alonso Carpio}
\email{mainak@fnal.gov, jose.dumler@nbi.ku.dk}
\author[0000-0002-2109-5315]{Mainak Mukhopadhyay}
\affiliation{Astrophysics Theory Department, Theory Division, Fermi National Accelerator Laboratory, Batavia, Illinois 60510, USA}
\altaffiliation{FERMILAB-PUB-26-0760-T}
\affiliation{Kavli Institute for Cosmological Physics, University of Chicago, Chicago, Illinois 60637, USA}
\affiliation{NSF-Simons AI Institute for the Sky (SkAI), 172 E. Chestnut St., Chicago, Illinois 60611, USA}
\affiliation{Department of Physics; Department of Astronomy \& Astrophysics; Institute for Gravitation and the Cosmos, The Pennsylvania State University, University Park, Pennsylvania 16802, USA}
\email{mainak@fnal.gov}
\author[0000-0003-0667-6557]{Jose Alonso Carpio}
\affiliation{Niels Bohr Institute, University of Copenhagen, Blegdamsvej 17, 2100 Copenhagen, Denmark}
\affiliation{Department of Physics \& Astronomy, Nevada Center for Astrophysics, University of Nevada, Las Vegas, NV 89514, USA}
\email{jose.dumler@nbi.ku.dk}
\author[0009-0000-8519-1077]{Heer Patel}
\affiliation{Department of Physics; Department of Astronomy \& Astrophysics; Institute for Gravitation and the Cosmos, The Pennsylvania State University, University Park, Pennsylvania 16802, USA}
\email{patelheer0805@gmail.com}
\author[0000-0001-7074-0539]{Ali Kheirandish}
\affiliation{Department of Physics \& Astronomy, Nevada Center for Astrophysics, University of Nevada, Las Vegas, NV 89514, USA}
\email{ali.kheirandish@unlv.edu}
\author[0000-0002-5358-5642]{Kohta Murase}
\affiliation{Department of Physics; Department of Astronomy \& Astrophysics; Institute for Gravitation and the Cosmos, The Pennsylvania State University, University Park, Pennsylvania 16802, USA}
\affiliation{Graduate School of Science, The University of Tokyo, 7-3-1 Hongo, Bunkyo, Tokyo 113-0033, Japan}
\affiliation{Center for Gravitational Physics and Quantum Information, Yukawa Institute for Theoretical Physics, Kyoto, Kyoto 606-8502 Japan.}
\email{murase@psu.edu}
\begin{abstract}
In high-energy neutrino astrophysics, the identification of high-energy neutrino sources is a central goal, and jet-quiet active galactic nuclei (AGNs) have emerged as promising candidates. For this purpose, we present a theory-guided, intrinsic X-ray selected sample of hidden neutrino AGN candidates consisting of $28$ X-ray bright AGNs (X-HiNAC), combining the {\em Swift}/BAT and BASS surveys with dedicated {\em NuSTAR} observations and spectral analyses. We compile their intrinsic X-ray fluxes, spectral properties, luminosity distances, and supermassive black hole masses, to provide a coherent sample for neutrino source searches. The sample is physically motivated but remains model independent given that the neutrino flux is proportional to the intrinsic X-ray flux. 
Using the magnetically powered corona model, we also demonstrate the prospects for individual-source and stacking searches with IceCube-Gen2 and KM3NeT. We identify promising individual targets and find that stacking the remaining sources, excluding NGC 1068, could yield a $5\sigma$ detection within $\sim 5-15$ years under favorable model assumptions at the current and upcoming neutrino observatories.
We highlight systematic uncertainties in the adopted source parameters, particularly the black hole masses, for which different measurement techniques can yield estimates differing by up to an order of magnitude. Such uncertainties can significantly affect the predicted neutrino emission, motivating more precise black hole mass measurements and improved broadband X-ray observations. Our findings motivate improved multiwavelength characterization of nearby AGNs and provide a guide for investigating their role as high-energy neutrino sources.
\end{abstract}
%
%
\section{Introduction}
Identifying the astrophysical sources of high-energy neutrinos remains one of the central goals of multimessenger astronomy. Since the discovery of the astrophysical neutrino flux by IceCube in 2013~\citep{IceCube:2013cdw,IceCube:2013low}, significant progress has been made toward identifying its origins. This includes evidence for neutrino emission from the Galactic plane~\citep{IceCube:2023ame,IceCube:2026plr}, 
and identification of neutrino emission from the nearby Seyfert galaxy NGC 1068~\citep{IceCube:2022der, IceCube:2024ayt, IceCube:2025lev}. More recently, IceCube has also reported evidence of neutrino emission from populations of X-ray-bright Seyfert galaxies~\citep{IceCube:2025gdd,Abbasi:2026oux}. Upcoming neutrino telescopes, including KM3NeT~\citep{KM3Net:2016zxf}, IceCube-Gen2~\citep{IceCube-Gen2:2020qha,Gen2_TDR}, and P-ONE~\citep{P-ONE:2020ljt}, will provide improved sensitivity for identifying individual neutrino sources and measuring their contributions to the all-sky astrophysical neutrino flux.

Among the proposed neutrino source populations, active galactic nuclei (AGNs) are particularly promising, especially following the evidence for neutrino emission from NGC 1068~\citep{IceCube:2022der,IceCube:2024ayt, IceCube:2025lev,IceCube:2025gdd}. Its neutrino emission, together with gamma-ray observations~\citep{Ajello:2023hkh} and upper limits at very high energies~\citep{MAGIC:2019fvw}, points to scenarios in which high-energy neutrinos originate from compact, gamma-ray obscured regions. In such environments, a large optical depth to $\gamma\gamma$ pair production ($\tau_{\gamma\gamma} \gg 1$) suppresses the escaping very-high-energy gamma-ray emission, with the absorbed energy reprocessed to lower energies, as required from all-sky neutrino and gamma-ray fluxes~\citep{Murase:2015xka}. The vicinity of supermassive black holes (SMBHs) provides a natural environment for such neutrino production, where magnetic dissipation and turbulence may accelerate nonthermal protons that produce high-energy neutrinos through hadronuclear ($pp$) and photohadronic ($p\gamma$) interactions~\citep{Murase:2019vdl,Inoue:2019yfs,Kheirandish:2021wkm,Murase:2022dog,Inoue:2022yak,Murase:2023ccp,Lemoine:2024roa,Fiorillo:2023dts,Murase:2026hrz,Carpio:2026xkf,LeBihan:2026rqq,Testagrossa:2026jcs}. The emission regions are constrained to be $\lesssim 15-70$ Schwarzschild radii ($R_{\rm S}$), which motivate studies about compact, X-ray-emitting coronae of AGNs~\citep{Murase:2022dog,Das_2024,Blanco:2023dfp,Das:2026rwt}.

The use of X-ray-selected AGN samples for neutrino searches has been explored in several previous studies. In particular, \citet{Murase:2019vdl} investigated neutrino production in AGN coronae and discussed the prospects for identifying nearby X-ray bright sources using the BASS sample.  They found more promising AGNs in the southern sky, which are good targets for upcoming neutrino detectors such as KM3NeT. Building on this approach, \citet{Kheirandish:2021wkm} investigated the expected neutrino emission from a sample of nearby X-ray-bright Seyfert galaxies and their detection prospects with current and future neutrino telescopes.
More recently, \citet{Murase:2023ccp} and \cite{Neronov:2023aks} showed the most promising neutrino-active galaxies, including NGC 4151. The IceCube Collaboration subsequently performed catalog-based searches for neutrino emission from Seyfert galaxies selected using {\em Swift}/BAT and BASS measurements~\citep{IceCube:2025gdd,Abbasi:2026oux}. 
These studies have established X-ray-selected AGN samples as promising targets for individual-source and stacking searches. However, the selection of candidates and their relative rankings depend on the adopted X-ray measurements, energy bands, treatment of intrinsic nuclear emission, and the exclusion of merger and jet activities.
 
In particular, earlier studies primarily used intrinsic $2-10$ keV X-ray fluxes to identify promising neutrino candidates, while more recent IceCube searches have also considered selection in harder X-ray bands, such as $20 - 50$ keV, which are generally less affected by photoelectric absorption. Nevertheless, even hard X-ray emission can be substantially attenuated in heavily obscured and Compton-thick AGNs, for which recovering the intrinsic nuclear continuum requires careful treatment of absorption, Compton scattering, and reflection~\citep{Marchesi:2018iie,Balokovic:2018bcm}. Furthermore, existing X-ray samples combine observations and spectral analyses performed under different assumptions, and updated broadband measurements can lead to significant revisions of the inferred intrinsic luminosities. Consequently, the ranking of promising neutrino candidates can depend significantly on the adopted X-ray band and spectral modeling. For example, NGC 4151 ranked fifth or sixth in the IceCube sky based on the intrinsic $2-10$ keV fluxes in BASS, but rose to the second brightest when considering the broader $2-195$ keV band. This highlights the importance of systematically comparing intrinsic X-ray fluxes across different energy ranges when identifying promising neutrino-source candidates.

Another important consideration is the uncertainty in the physical properties of the central SMBHs. In particular, SMBH mass estimates obtained using different measurement techniques can differ substantially~\citep{10.1093/mnras/stad3974}. Since the SMBH mass enters the Eddington ratio and influences the physical conditions governing neutrino production, these uncertainties can significantly affect the predicted neutrino spectra and detection prospects. A consistently characterized sample incorporating updated intrinsic X-ray measurements, spectral properties, luminosity distances, and SMBH mass estimates is therefore valuable for systematic comparisons among candidate sources and for guiding future neutrino searches.

Motivated by these considerations, we construct a sample of X-ray bright AGNs for high-energy neutrino searches with current and next-generation neutrino telescopes. We refer to this sample as \textbf{X-HiNAC} (intrinsic X-ray selected Hidden Neutrino Active galaxy Candidates). We combine the BASS and {\em Swift}/BAT surveys with dedicated {\em NuSTAR} observations and broadband spectral analyses, and apply a set of selection and validation criteria to obtain a final sample of $28$ AGNs. For these sources, we compile their intrinsic X-ray fluxes in the $2 - 10$ keV and $2 - 195$ keV bands, spectral properties, luminosity distances, and SMBH masses. The two energy bands provide complementary measures of the intrinsic emission and allow us to examine how the ranking of promising neutrino candidates depends on the adopted X-ray energy range. Importantly, the sample is constructed independently of the detailed neutrino production model used for our forecasts and provides a reference sample for scenarios in which the neutrino flux approximately scales with the intrinsic X-ray flux ($F_\nu \propto F_{X,\rm intr}$).

As a representative application, we adopt the magnetically powered corona model~\citep{Murase:2019vdl,Murase:2026hrz} to calculate benchmark fluxes of the expected neutrino emission under three model assumptions. We then estimate the corresponding neutrino event rates at IceCube-Gen2 and KM3NeT, accounting for their effective areas, angular resolutions, and atmospheric neutrino backgrounds, and investigate the prospects for individual-source and stacking searches. We also evaluate the combined neutrino flux from our sample and compare it with the diffuse astrophysical neutrino flux measured by IceCube. Beyond these detection forecasts, our study highlights the observational uncertainties that currently limit reliable neutrino predictions and provides a reference sample for future neutrino searches.

The paper is organized as follows. In Section~\ref{sec:nu_agn} we review the mechanisms of high-energy neutrino production in X-ray-bright AGNs and describe the magnetically powered corona model adopted in this work. We present the construction of our AGN sample, including the source selection criteria and the adopted observational properties in Section~\ref{sec:cat}. The predicted neutrino spectra and detection prospects for IceCube-Gen2 and KM3NeT, including individual source and stacking searches, are presented in Section~\ref{sec:res}. Finally, in Section~\ref{sec:concl}, we summarize our main findings and discuss the implications of our work.
\section{High-energy neutrino production in Jet-Quiet AGNs}
\label{sec:nu_agn}
Jet-quiet AGNs are promising sources of high-energy neutrinos, particularly if cosmic rays (CRs) are accelerated in compact regions surrounding the central SMBH. In such environments, relativistic protons can interact with the ambient gas and radiation fields through hadronuclear ($pp$) and photohadronic ($p\gamma$) processes, producing high-energy neutrinos through the decay of charged pions. The dense radiation fields can also efficiently attenuate the accompanying high-energy gamma rays, making these regions potential gamma-ray obscured neutrino sources~\citep[see a review,][and references therein]{Murase:2022feu}.

A representative scenario is the magnetically powered corona model~\citep{Murase:2019vdl}, in which magnetic fields generated in the accretion disk emerge in the corona and dissipate through magnetic reconnections and turbulence. The released magnetic energy heats the coronal plasma, whose thermal electrons produce the observed X-ray emission through inverse Compton scattering, while nonthermal protons are accelerated through Fermi processes. The resulting neutrino spectrum depends on the competition between particle acceleration, hadronic interactions, energy losses, and escape from the corona.

An important feature of such scenarios is the connection between the intrinsic X-ray luminosity and the neutrino luminosity~\citep{Murase:2019vdl,Kheirandish:2021wkm,Murase:2023ccp}. Given that the system is nearly calorimetric, the neutrino luminosity approximately scales as the cosmic-ray power, which enables us to expect
\be
\label{eq:cr_xray_relation}
L_{\nu} \propto L_X^{\beta_\nu}\,,
\ee
where $\beta_\nu$ depends on the underlying coronal properties. Detailed calculations of the magnetically powered corona model yield $\beta_\nu \sim 0.8 - 1.1$ for the representative models considered here~\citep{Murase:2026hrz}, and the above relationship is consistent with existing data within large uncertainties~\citep{Carpio:2026xkf,Das:2026aed}. It is important to note that the inferred normalization depends on the assumed neutrino spectral shape and the energy range over which the neutrino luminosity is evaluated~\citep{Ambrosone:2024zrf,Carpio:2026xkf}. For example,~\cite{Kun:2024meq} obtain significantly higher values of $L_\nu$ by extrapolating the assumed power-law spectra neutrino spectra to $300$ GeV, whereas \citet{Carpio:2026xkf} obtained more conservative values through the corona model. 

The resulting approximately linear relation, $F_\nu \propto F_{X,\rm intr}$ motivates the use of intrinsic X-ray fluxes to identify promising neutrino-source candidates, and it does not depend on details of the magnetically powered corona model. 

The physical conditions governing neutrino production also depend on the properties of the accretion flow. As pointed out by \citet{Murase:2026hrz}, one of the relevant quantities in modeling neutrino spectra is the Eddington ratio,
\be
\label{eq:eddington_ratio}
\lambda_{\rm Edd} \equiv \frac{L_{\rm bol}}{L_{\rm Edd}}\,,
\ee
where $L_{\rm Edd} = \big( 4\pi G M_{\rm BH}m_p c \big)/\sigma_T$, $L_{\rm bol}$ is the bolometric luminosity, and $M_{\rm BH}$ is the SMBH mass. The Eddington ratio characterizes the accretion state and can influence the coronal size and its radiation environments, thereby affecting the neutrino spectrum. Consequently, even for a given intrinsic X-ray luminosity, differences in $\lambda_{\rm Edd}$ can lead to variations in the predicted neutrino emission. Since $\lambda_{\rm Edd} \propto L_X/M_{\rm BH}$, uncertainties in SMBH mass measurements can therefore introduce additional uncertainties in the neutrino predictions.

\subsection{Relevance of Model-Dependent Searches}
It is important to note that realistic neutrino spectra do not have to be power laws that are widely used for the purpose of fitting. Indeed, many astrophysical models predict convex spectra~\cite[e.g.,][]{Murase:2015xka,Murase:2019vdl}, which is also supported by evidence for the spectral break around $30$ TeV~\citep{IceCube:2025tgp}. Neutrino spectra that are not described by simple power laws can significantly alter detection prospects and therefore it is crucial to consider model-dependent neutrino spectra for the purpose of neutrino source searches.

To illustrate the detection prospects of our sample with model-dependent neutrino spectra, we adopt the magnetically powered corona model developed in~\citet{Murase:2019vdl,Kheirandish:2021wkm,Murase:2026hrz}. Following the numerical treatment of~\citet{Murase:2026hrz}, we employ the \texttt{AMES} (Astrophysical Multimessenger Emission Simulator) framework to calculate the neutrino spectra under three representative model assumptions. \textbf{Model A} corresponds to one of the original representative cases in the corona model by~\citet{Murase:2019vdl}, while \textbf{Model B} and \textbf{Model C} explore different assumptions for the AGN population and turbulent acceleration. 

For demonstrative purposes, we choose these three sets of physical parameters that are motivated by modeling the all-sky astrophysical neutrino spectrum. The calculation accounts for turbulent acceleration, hadronic interactions, and the relevant cooling and escape processes. We use the source specific intrinsic X-ray luminosities, spectral properties, and SMBH masses compiled in Section~\ref{sec:cat} as inputs to these calculations. Importantly, in this work, we do not independently optimize the coronal parameters for each AGN or incorporate their intrinsic dispersion from one source to another.

Specifically, Models A and B adopt $q=5/3$ with $\eta_{\rm tur} = 10$ and $20$, respectively, while Model C assumes $q = 2$ (hard-sphere), $\eta_{\rm tur} = 15$, and stronger magnetic fields, where $q$ characterizes the energy dependence of the diffusion coefficient and $\eta_{\rm tur}$ parameterizes the acceleration efficiency, with smaller values corresponding to more efficient particle acceleration. For the calculation of the neutrino spectra, we adopt a dimensionless coronal radius $\mathcal{R}\equiv R_{\rm c}/R_{\rm S}$, where $R_{\rm c}$ is the characteristic coronal radius and $R_{\rm S} = 2GM_{\rm BH}/c^2$ is the Schwarzschild radius. For Models A and B, we assume $\mathcal{R} = 3$ for $\lambda_{\rm Edd} \geq 0.3$ and $\mathcal{R} = 10$ for $\lambda_{\rm Edd} < 0.3$, while for Model C, we fix $\mathcal{R} = 10$ irrespective of the Eddington ratio. These representative choices allow us to investigate the neutrino emission from our AGN sample under different coronal conditions, but we note that these are adopted just as representative benchmarks. 

\section{Construction of the X-H\MakeLowercase{i}NAC sample}
\label{sec:cat}
\begin{deluxetable*}{llcccccccc}
\tabletypesize{\scriptsize}
\tablewidth{\textwidth}
\tablecaption{\label{tab:swift_bat_sample}X-HiNAC: Selected AGN sample along with their properties and corresponding tiers within the sample in the $2 - 10\ {\rm keV}$ and $2 - 195\ {\rm keV}$ bands based on the intrinsic X-ray flux. The rankings in the respective bands are denoted by ${\rm R}_{2-10}$ and ${\rm R}_{2-195}$. The Sky column denotes the celestial hemisphere based on the source declination: 'N' corresponds to the Northern sky ($\delta \geq 5^\circ$), 'S' corresponds to the Southern sky ($\delta \leq - 5^\circ$), while 'H' denotes the Horizon ($|\delta| < 5^\circ$). This classification provides a convenient indication of the sources most relevant to IceCube and KM3NeT, respectively. The intrinsic X-ray fluxes
$F_{X,\mathrm{intr}}^{2 - 10\ \mathrm{keV}}$ and
$F_{X,\mathrm{intr}}^{2 - 195\ \mathrm{keV}}$ are given in units of
$10^{-12}\ \mathrm{erg\ cm^{-2} s^{-1}}$. The adopted X-ray spectral index is given by $\Gamma_X$, the hydrogen-equivalent absorbing column density $N_{\rm H}$ is given in $\mathrm{cm}^{-2}$, the black hole mass $M_{\rm BH}$ is given in units of $10^{6}\ M_\odot$, the luminosity distance $d_L$ is given in Mpc. 
}
\tablehead{
\colhead{Name} &
\colhead{Sky} &
\colhead{$F_{X,\mathrm{intr}}^{2\text{--}10\,\mathrm{keV}}$} &
\colhead{$F_{X,\mathrm{intr}}^{2\text{--}195\,\mathrm{keV}}$} &
\colhead{$\mathrm{Tier}_{2\text{--}10}/\mathrm{Tier}_{2\text{--}195}$} &
\colhead{$\mathrm{R}_{2\text{--}10}/\mathrm{R}_{2 - 195}$} &
\colhead{$\Gamma_X$} &
\colhead{$\log_{10} (N_{\rm H}/{\rm cm}^{-2})$} &
\colhead{$M_{\rm BH}$} &
\colhead{$d_L$}
}
\startdata
NGC 1068 &
H &
$2821$ &
$6973$ &
I/I &
$1/1$ &
2.10 & 25.0 &
$6.0 \pm 0.2$\tablenotemark{b, $\delta$} &
10.7
\\
Circinus Galaxy &
S &
$1500$ &
$4230$ &
I/I &
$2/2$ &
2.30 & 24.4 &
$1.7\pm0.3$\tablenotemark{a} &
4.20
\\
NGC 4945 &
S &
132 &
614 &
III/II &
$7/3$ &
1.61 & 24.5 &
$1.3 \pm 0.7$\tablenotemark{c, $\delta$} &
3.55
\\
NGC 4151\tablenotemark{*} &
N &
166 &
600 &
II/II &
$6/4$ &
1.85 & 22.7 &
$17^{+4.8}_{-3.4}$\tablenotemark{d} &
14.4
\\
IC 4329A\tablenotemark{*} &
S &
171 (95.7) &
481 &
II/II &
$5/5$ &
2.01 & 21.5 &
$105^{+24}_{-29}$\tablenotemark{e} &
70.4
\\
NGC 2110\tablenotemark{*} &
S &
116 &
456 &
III/II &
$8/6$ &
1.80 & 22.9 &
$25^{+21}_{-15}$\tablenotemark{f} &
34.1
\\
NGC 5506\tablenotemark{*} &
H &
187 (69.0) &
455 &
II/II &
$4/7$ &
2.11 & 22.4 &
17\tablenotemark{g} &
27.1
\\
MCG-05-23-16\tablenotemark{*} &
S &
197 (120) &
440 &
II/II &
$3/8$ &
2.18 & 22.2 &
45\tablenotemark{g} &
37.2
\\
NGC 4507\tablenotemark{*} &
S &
82.7 &
299 &
III/III &
$15/9$ &
1.85 & 24.0 &
45\tablenotemark{h} &
51.7
\\
NGC 3783\tablenotemark{*} &
S &
97.2 &
285 &
III/III &
$12/10$ &
1.98 & 20.5 &
$25.4^{+9.0}_{-7.2}$\tablenotemark{i} &
42.5
\\
NGC 7172 &
S &
83.8 &
279 &
III/III &
$14/11$ &
1.87 & 24.1 &
$10 \pm 3.5$\tablenotemark{j} &
37.5
\\
UGC 3374\tablenotemark{*} &
N &
97.3 &
250 &
III/III &
$11/12$ &
2.07 & 20.5 &
$52^{+16}_{-2.3}$\tablenotemark{k} &
90.5
\\
GRS 1734-292\tablenotemark{*} &
S &
110 (73.6) &
236 &
III/III &
$9/13$ &
2.21 & 21.92 &
$316$\tablenotemark{l} &
93.2
\\
Mrk 348\tablenotemark{*} &
N &
61.9 &
232 &
IV/III &
$22/14$ &
1.83 & 23.1 &
16\tablenotemark{m} &
65.9
\\
Mrk 3 &
N &
49.1 &
230 &
IV/III &
$27/15$ &
1.60 & 23.9 &
$180^{+110}_{-70}$\tablenotemark{f} &
58.5
\\
ESO 103-35\tablenotemark{*} &
S &
102 (43.0) &
229 &
III/III &
$10/16$ &
2.18 & 23.3 &
32\tablenotemark{n} &
58.4
\\
ESO 506-27 &
S &
59 &
197 &
IV/IV &
$23/17$ &
1.93 & 23.9 &
$398$\tablenotemark{n} &
109
\\
NGC 3227\tablenotemark{*} &
N &
69 &
196 &
IV/IV &
$18/18$ &
2.00 & 21.0 &
$11^{+2}_{-3}$\tablenotemark{o} &
17.0
\\
IRAS 05078+1626\tablenotemark{*} &
N &
79.8 &
195 &
IV/IV &
$16/19$ &
2.11 & 21.1 &
$35 \pm 5$\tablenotemark{p} &
78.8
\\
Mrk 509\tablenotemark{*} &
S &
75.9 (139) &
193 &
IV/IV &
$17/20$ &
2.08 & 20.0 &
$102\pm 45$\tablenotemark{e} &
153
\\
4U1344-60\tablenotemark{*} &
S &
66.4 (47.0) &
192 &
IV/IV &
$19/21$ &
1.99 & 22.1 &
$20^{+20}_{-10}$\tablenotemark{q} &
56.6
\\
NGC 7479 &
N &
56.8 &
187 &
IV/IV &
$24/22$ &
1.85 & 24.8 &
$12$\tablenotemark{r} &
34.2
\\
MR 2251-178\tablenotemark{*} &
S &
66.1 &
183 &
IV/IV &
$20/23$ &
2.02 & 20.0 &
$213 \pm 53$\tablenotemark{s, $\delta$} &
292
\\
NGC 3281 &
S &
52.0 (8.95) &
182 &
IV/IV &
$26/24$ &
2.18 & 24.0 &
170\tablenotemark{g} &
46.2
\\
NGC 6300\tablenotemark{*} &
S &
64.3 &
180 &
IV/IV &
$21/25$ &
2.01 & 23.3 &
$48$\tablenotemark{t, $\delta$} &
16.1
\\
Mrk 926\tablenotemark{*} &
S &
53.6 &
173 &
IV/IV &
$25/26$ &
1.92 & 20.0 &
$110 \pm 20$\tablenotemark{u} &
211
\\
MCG-6-30-15\tablenotemark{*} &
S &
95.2 &
172 &
III/IV &
$13/27$ &
2.38 & 20.9 &
$58 \pm 18$\tablenotemark{v, $\delta$} &
33.7
\\
NGC 5728 &
S &
32.6 (70.1) &
162 &
IV/IV &
$28/28$ &
1.55 & 24.1 &
34\tablenotemark{w} &
40.3
\enddata
\tablecomments{The black-hole-mass references are
$^{a}$\cite{2003ApJ...590..162G},
$^{b}$\cite{Lodato:2002cv,2024ApJ...975L...9G,2026ApJ..1000...78M},
$^{c}$\cite{Greenhill:1996gz},
$^{d}$\cite{2022ApJ...934..168B},
$^{e}$\cite{2024ApJ...974...86L},
$^{f}$\cite{2015ApJ...802...98M},
$^{g}$\cite{Serafinelli_2024},
$^{h}$\cite{Marinucci_2013},
$^{i}$\cite{GRAVITY:2021ory},
$^{j}$\cite{Shuvo:2022cxw},
$^{k}$\cite{2025ApJ...979..131F},
$^{l}$\cite{Tortosa:2016loq},
$^{m}$\cite{Woo_2002},
$^{n}$\cite{Kawamuro_2016},
$^{o}$\cite{bentz2023velocityresolvedreverberationmappingngc},
$^{p}$\cite{ponti_etal},
$^{q}$\cite{2025ApJ...994..151G},
$^{r}$\cite{Pizzetti_2022},
$^{s}$\cite{Lira_2011},
$^{t}$\cite{2012AstL...38..475K},
$^{u}$\cite{10.1093/mnras/stac2953},
$^{v}$\cite{das2025stellardynamicalmasscentral}, and
$^{w}$\cite{2019ApJ...870...37D}.
}
\tablenotetext{*}{The intrinsic $2-10$ keV flux was inferred from the observed $14-195$ keV flux.}
\tablenotetext{\delta}{The candidates for which $M_{\rm BH}$ was rescaled with $d_L$.}
\end{deluxetable*}

In this section, we describe the construction of the X-HiNAC sample, including the adopted observational data, source selection criteria, and characterization of the intrinsic X-ray emission. The main aim of such a sample is to report the most appropriate values for the (observed) intrinsic X-ray flux ($F_{X,\rm intr}$) and the SMBH mass $M_{\rm BH}$, which primarily depend on the luminosity distance $d_L$, spectral index $\Gamma_X$, and the hydrogen-equivalent absorbing column density $N_H$. In this work, we try to perform a systematic study across the major X-ray surveys and the literature to narrow down on these parameters. We then provide a distinct ranking based on $F_{X,\rm intr}$ in the $2- 10 $ keV and the $2- 195$ keV X-ray bands for each AGN candidate. As a way to select out the most promising candidates, we place a flux cut corresponding to the $5 \sigma$ discovery potential of IceCube-Gen2. The final X-HiNAC sample with the relevant parameters is presented in Table~\ref{tab:swift_bat_sample}, while additional details about the sources in the sample are presented in Appendix~\ref{appsec:details_cat} and Table~\ref{tab:swift_bat_coordinates}.

We used four surveys and samples for this project.
\begin{itemize}
\item \textbf{BAT AGN Spectroscopic Survey (BASS):} the baseline sample of {\em Swift}/BAT-selected AGNs and their broadband X-ray spectral properties~\citep{Ricci:2017dhj,Koss:2017ilk,Koss:2022lkb,Koss:2022qwh}.
\item \textbf{{\em Swift}/BAT 157-month all-sky hard-X-ray survey:} the most recent BAT measurements of the $14 - 195\ {\rm keV}$ emission~\citep{Lien:2025zzo,SwiftBAT157Catalog}.
\item \textbf{BASS torus-model analysis:} dedicated spectral fits of heavily obscured BASS AGNs using the torus model of~\citet{Ricci:2017dhj}.
\item \textbf{Nuclear Spectroscopic Telescope Array ({\em NuSTAR}) observations of candidate Compton-thick AGNs:} broadband spectral fits using the XClumpy torus model~\citep{Tanimoto:2022wrq}.
\end{itemize}
\subsection{Baseline BASS sample}
We begin with the 70-month BASS sample, which contains 838 AGNs. Of these, 105 are classified as blazars and are excluded because their X-ray emission may be dominated by relativistically beamed jet emission. This leaves a baseline sample of 733 non-blazar AGNs.

For each source, we extract the $z$, $\Gamma_X$, $E_{X,\rm cut}$, $N_H$, and the intrinsic fluxes and luminosities in the $2$–$10\ {\rm keV}$ and $14$–$195\ {\rm keV}$ bands. These quantities provide the initial set of X-ray properties adopted in our analysis. Sources whose parameters are taken exclusively from BASS are assigned the sample identifier \textbf{1}.
\subsection{Swift/{\rm BAT} 157-month sample}
We next consider the {\em Swift}/BAT 157-month all-sky hard-X-ray survey, which contains 1888 detected sources. We select 1066 objects classified as Seyfert 1, Seyfert 2, or other AGNs, consisting of 514 Seyfert 1 galaxies, 462 Seyfert 2 galaxies, and 90 sources classified as other AGNs. These objects form the primary hard X-ray selected sample considered in this work. For each source, we retain $z$, $\Gamma_X$, and the reported flux and luminosity in the $14$–$195\ {\rm keV}$ band. For this sample, we ignore the exponential part of the spectrum (see Equation~\ref{eq:photon_powerlaw}), implying $E_{X,\rm cut}$ larger than the $195$ keV.

The 157-month sample is cross-matched with the baseline BASS sample. For the 717 sources common to both samples, we adopt the more recent $14$–$195\ {\rm keV}$ measurements from the 157-month sample while retaining the complementary information available from BASS. These sources are assigned the sample identifier \textbf{12}. The 16 objects present only in the BASS sample retain identifier \textbf{1}, whereas the 349 objects found only in the 157-month sample are assigned identifier \textbf{2}.

Furthermore, for sources that are securely Compton-thin, the attenuation in the BAT energy range is relatively weak, and we therefore use the reported hard X-ray flux as an approximation to the intrinsic $14$–$195\ {\rm keV}$ flux. For heavily obscured sources, however, this approximation can become inaccurate, and we instead prioritize the results of dedicated torus-model analyses described below.
\subsection{Torus-model analysis of heavily obscured BASS AGNs}
Simple absorbed power-law models become increasingly unreliable at large column densities because Compton scattering, fluorescent-line emission, and reprocessed continuum emission modify the observed spectrum. We therefore supplement the baseline sample with the dedicated BASS analysis of 75 heavily obscured sources fitted using the spherical-toroidal model of~\citet{Brightman:2011}.

The torus model provides improved constraints on the intrinsic continuum and obscuring column, particularly for sources with $N_{\rm H}\gtrsim10^{24}\ {\rm cm}^{-2}$. When a source is included in this analysis, we replace the baseline photon index ($\Gamma_X$) and column density ($N_H$) with the corresponding torus-model values. Because these objects are already contained in the BASS sample, no additional sources are introduced at this stage. Similar to the BAT-157 month sample, we used a power law X-ray spectrum which did not require us to choose $E_{X,\rm cut}$ for the candidates contained in this sample. Sources incorporating information from BASS, the 157-month survey, and the dedicated torus analysis are assigned the sample identifier \textbf{123}.
%
\subsection{NuSTAR sample of candidate Compton-thick AGNs}
Finally, we incorporate the broadband spectral analysis of 52 candidate Compton-thick AGNs selected from the {\em Swift}/BAT survey and observed with {\em NuSTAR}~\citep{Tanimoto:2022wrq}. The analysis combines {\em NuSTAR} observations with available data from Chandra, XMM-Newton, {\em Swift}/XRT, and Suzaku and uses the XClumpy model to account for reprocessing by an inhomogeneous circumnuclear torus. Sources whose properties are adopted from this sample are assigned the sample identifier \textbf{4}. Accordingly, sources that also appear in the previous samples are assigned the corresponding combined sample identifier (e.g., \textbf{1234})

For these sources, we compile $z$, $\Gamma_X$, $N_H$, observed fluxes in the $2$–$10\ {\rm keV}$ and $10$–$50\ {\rm keV}$ bands, and the corresponding absorption-corrected luminosities. For all sources in this sample, the authors adopted $E_{X,\rm cut} = 370$ keV and we use the same for the sources that we tabulate from this sample. These measurements are particularly important for estimating the intrinsic emission of highly obscured systems, for which the directly observed soft X-ray flux may substantially underestimate the nuclear luminosity. When multiple measurements are available for the same source, we prioritize the broadband {\em NuSTAR}/XClumpy results for the photon index, column density, and intrinsic $2$–$10\ {\rm keV}$ luminosity. 

The successive cross-matching and parameter updates described above produce a unified sample in which the most recent BAT measurements are used for the hard X-ray emission, while dedicated torus-model results are adopted for heavily obscured and candidate Compton-thick sources. The procedure used to convert between luminosities, fluxes, and energy bands, as well as the prescription for assigning SMBH masses, is described below.
\subsection{Unifying the sample and selecting promising candidates}
The unification of the samples resulted in $1083$ sources to begin with. The samples described above report X-ray fluxes and luminosities over different energy ranges. To compare the sources uniformly, we convert the available measurements to two common energy bands, $2$–$10\ {\rm keV}$ and $2$–$195\ {\rm keV}$. We model the intrinsic X-ray spectrum of each source as a cutoff power law
\be
\label{eq:photon_powerlaw}
E F_E = \mathcal{N}_X E^{1-\Gamma_X} \exp\left(-\frac{E}{E_{X,\rm cut}}\right)\,,
\ee
where $\mathcal{N}_X$ is the spectral normalization, $\Gamma_X$ is the adopted photon index, and $E_{X,\rm cut}$ is the high-energy cutoff. The corresponding energy flux in an interval $(E_1,E_2)$ is given by
\be
\label{eq:xray_band_flux}
F_{X,\rm intr}^{[E_1,E_2]} = \mathcal{N}_X \int_{E_1}^{E_2}dE\ E^{1-\Gamma_X} \exp\left(-\frac{E}{E_{X,\rm cut}}\right)\,.
\ee
Whenever flux measurements are available in two different energy bands, we independently infer a normalization from each measurement. For a reported flux $F_{X,{\rm intr},i}$ in the interval $(E_{i,1},E_{i,2})$, the corresponding normalization can be defined as
\be
\mathcal{N}_{X,i} = \frac{F_{X,{\rm intr},i}}{\displaystyle \int_{E_{i,1}}^{E_{i,2}}dE\ E^{1-\Gamma_X} \exp\left(-\frac{E}{E_{X,\rm cut}}\right)}\,.
\ee
This provides a consistency check between the spectral measurements reported in the two bands. We ensure that for the sources the independently inferred normalizations agree to within a factor of $\sqrt{10}$, that is, $\left|\log_{10}\left(\mathcal{N}_{X,1}/\mathcal{N}_{X,2}\right)\right|\leq 0.5$. We then adopt an average normalization $\mathcal{N}_{X,\rm avg}=(\mathcal{N}_{X,1}+\mathcal{N}_{X,2})/2$ and use Equation~\eqref{eq:xray_band_flux} to determine the fluxes in the common $2$–$10\ {\rm keV}$ and $2$–$195\ {\rm keV}$ intervals. This procedure allows us to combine, for example, the $2$–$10\ {\rm keV}$ measurements commonly reported by soft-X-ray instruments with the $14$–$195\ {\rm keV}$ measurements from {\em Swift}/BAT. It also reduces the sensitivity of the inferred broadband flux to any one individual measurement.

The quantity relevant to our analysis is the intrinsic, or absorption-corrected, energy flux at Earth. Some of the samples provide this quantity directly, whereas others report an observed flux together with an absorption-corrected luminosity. In the latter case, we infer the intrinsic flux from the reported luminosity according to
\be
\label{eq:intr_flux_lum}
F_{X,\rm intr}^{[E_1,E_2]} = 
\frac{L_X^{[E_1,E_2]}}{4\pi d_L^2}
(1+z)^{2-\Gamma_X}\,.
\ee
The final factor accounts for the $K$-correction for a power-law spectrum when relating luminosities and fluxes defined over the corresponding rest-frame and observer-frame energy intervals. For heavily obscured sources, we preferentially use the absorption-corrected luminosities obtained from the dedicated torus-model analyses rather than correcting the observed flux using a simple absorbed power-law model.

After placing all sources on a common footing, we rank the AGNs independently according to their intrinsic fluxes in the $2$–$10\ {\rm keV}$ and $2$–$195\ {\rm keV}$ bands. The former facilitates comparison with the commonly reported intrinsic coronal luminosity, whereas the latter provides an estimate of the total X-ray power over the broadband energy range relevant to our neutrino-production model.

To identify the most promising neutrino-source candidates, we impose a minimum broadband X-ray flux $F_{X,\rm lim}$ such that we retain only sources satisfying
\be
\label{eq:intr_xray_flux_cut}
F_{X,\rm intr}^{2 - 195\ \rm keV} \geq F_{X,\rm lim} = 10^{-7}\ {\rm GeV\ cm^{-2}\ s^{-1}}\,,
\ee 
where we choose $F_{X,\rm lim}$ motivated by the projected $5\sigma$ point-source discovery potential of IceCube-Gen2~\citep{Murase:2026hrz}. The application of this threshold is intended as a practical down-selection criterion rather than a direct prediction of the neutrino flux, since the precise relation between the X-ray and neutrino luminosities depends on the properties of the corona and on the efficiency of CR acceleration and interaction. Following this implementation, we were left with $39$ sources. The resulting candidates and their tiers or rankings in the two energy bands are presented in Table~\ref{tab:swift_bat_sample}. Source-specific modifications and cases requiring additional treatment are discussed in the following subsection.
\subsection{A note on adopted black hole masses and intrinsic dispersion}
In this section we discuss the SMBH masses that we adopt for our work and in particular, the intrinsic dispersion associated with SMBH masses depending on the methods adopted to compute them.

SMBH masses in the literature are obtained using a variety of techniques, including stellar or gas dynamical modeling~\citep[e.g.,][]{das2025stellardynamicalmasscentral}, water-megamaser kinematics~\citep[e.g.,][]{2003ApJ...590..162G}, reverberation mapping~\citep[e.g.,][]{2024ApJ...976..176F,bentz2023velocityresolvedreverberationmappingngc,2024ApJ...974...86L}, interferometric reverberation mapping~\citep[e.g.,][]{GRAVITY:2021ory}, single-epoch virial estimators based on the broad-line
radius luminosity relation~\citep[e.g.,][]{Lira_2011}, and empirical host-galaxy scaling relations such as the $M_{\rm BH}$--$\sigma_\star$ and bulge-luminosity relations~\citep[e.g.,][]{Woo_2002,2012AstL...38..475K}.

Since these methods do not all share the same dependence on the assumed
distance, we examined each measurement individually before adopting it.
For measurements whose inferred physical scale depends explicitly on the
assumed distance (e.g., stellar, gas, or maser dynamics, and
single-epoch virial estimates), we recomputed the SMBH mass using
the luminosity distance ($d_L$) adopted in this work. For reverberation-mapping measurements and empirical $M_{\rm BH}$--$\sigma_\star$ relations, no rescaling was applied because the published masses are effectively independent of the assumed luminosity distance. Whenever the adopted mass was obtained from a heterogeneous compilation or from methods with more complicated distance dependence, we retained the published values.

Furthermore, the uncertainty in $M_{\rm BH}$, and in particular the systematic dispersion among estimates obtained using different methods, plays an important role in the resulting neutrino spectra and expected event distributions in neutrino detectors. This is primarily because $\lambda_{\rm Edd}$ depends on $M_{\rm BH}$ (see Section~\ref{sec:nu_agn}). This is particularly relevant for our calculation since the physical properties of the accretion flow, and hence the predicted neutrino emission, depend on $\lambda_{\rm Edd}$. The uncertainty associated with $M_{\rm BH}$ should therefore not be interpreted solely in terms of the statistical uncertainty quoted for an individual measurement, since for some sources, the spread among measurements obtained using different methods can be substantially larger. Below are a few (non-exhaustive) examples.
\begin{itemize}
\item NGC 4151: We adopt $M_{\rm BH} = 1.66^{+0.48}_{-0.34} \times 10^7 M_\odot$ from the velocity resolved reverberation mapping analysis of~\cite{2022ApJ...934..168B}, which is also broadly consistent with the mass estimate compiled by~\cite{Kormendy:2013dxa}. However, more recent reverberation-mapping observations by~\cite{2024ApJ...976..176F} yield appreciably larger values ($\sim (3.5 \pm 0.2) \times 10^{7} M_\odot$), with the inferred mass also varying between observing seasons. NGC4151 therefore provides a useful example of the systematic dispersion among $M_{\rm BH}$ estimates, which can exceed the statistical uncertainty associated with any individual measurement and translate directly into a corresponding uncertainty in the inferred $\lambda_{\rm Edd}$.
\item GRS 1734-292: We adopt $M_{\rm BH} \simeq 3.16 \times 10^8\ M_\odot$ from~\citet{Tortosa:2016loq}, estimated using the broad H$\alpha$ emission line and intrinsic X-ray luminosity. In particular, the authors measure ${\rm FWHM}({\rm H}\alpha) = 4940 \pm 50\ {\rm km\ s^{-1}}$ and to obtain $M_{\rm BH}$. However, the calibration has an intrinsic scatter of approximately $0.5$ dex, corresponding to a factor of $\sim 3.2$ in $M_{\rm BH}$, or a characteristic range of $\sim 10^8 - 10^9\ M_\odot$ around the adopted value.
\item NGC 7172: We adopt $M_{\rm BH} = (1.03 \pm 0.35) \times 10^7 M_\odot$ from~\citet{Shuvo:2022cxw}, obtained using an independent X-ray spectral scaling method. The measured stellar velocity dispersion gives $M_{\rm BH} \simeq 5.5\times10^7 M_\odot$~\citep{2012ApJ...748..130M} inferred from $\sigma_\star = 160 \pm 9\ {\rm km\ s^{-1}}$, using the $M_{\rm BH}$–$\sigma_\star$ relations. However, the intrinsic scatter associated with this relation is substantial. The value adopted by~\citet{Kawamuro_2016}, $M_{\rm BH} \simeq 10^8 M_\odot$, is larger. The resulting spread of nearly an order of magnitude further illustrates the systematic uncertainty associated with individual SMBH mass estimates and the corresponding uncertainty in $\lambda_{\rm Edd}$.
\item IRAS 05078+1626: We adopt $M_{\rm BH} \simeq (3.5 \pm 0.5) \times 10^7 M_\odot$ from~\citet{ponti_etal} and classify the mass as an estimated value. The systematic uncertainty is further illustrated by the substantial spread among values adopted in the literature. For example,~\citet{Laha:2014faa} adopt $M_{\rm BH}\simeq 7.2 \times10^7 M_\odot$, whereas the recent X-WING compilation adopts $7.6 \times 10^6 M_\odot$~\citep{Yamada:2024lss}, spanning nearly an order of magnitude.
\item NGC 5728: We adopt $M_{\rm BH} = 3.4\times10^7\,M_\odot$ from~\citet{2019ApJ...870...37D}. However, substantially different SMBH mass estimates have been obtained using other methods. In particular~\citet{10.1093/mnras/stad3974} infer $M_{\rm BH} = 1.77 \times 10^8 M_\odot$ from the measured stellar velocity dispersion, $\sigma_\star=176\ {\rm km\ s^{-1}}$, using the $M_{\rm BH} - \sigma_\star$ relation, while an independent X-ray scaling method considered in the same work gives $M_{\rm BH} = 9.93^{+7.03}_{-3.41}\times10^6 M_\odot$. NGC 5728 also hosts water megamaser emission, however, VLBI observations reveal complex maser kinematics rather than a simple Keplerian disk, preventing a robust dynamical determination of $M_{\rm BH}$~\citep{2019arXiv191014314K}.
\end{itemize}
Similar method-dependent differences occur for several of the AGNs in our sample. We therefore emphasize that the $M_{\rm BH}$ values in Table~\ref{tab:swift_bat_sample} should be regarded as a consistent set of fiducial estimates rather than uniquely determined SMBH masses. Our fiducial results use the values listed in the table, while the impact of the dispersion in the published $M_{\rm BH}$ measurements can be assessed by varying the masses, and consequently $\lambda_{\rm Edd}$, over the range allowed by alternative measurements. This provides a direct estimate of the corresponding systematic uncertainty in the predicted neutrino spectra and event distributions.
\subsection{Comparison with {\em NuSTAR} intrinsic X-ray luminosities}
\label{sec:nustar_comparison}
As discussed before, wherever available, we prioritize dedicated {\em NuSTAR} spectral analyses in constructing our sample. Such observations are particularly valuable for constraining the intrinsic
X-ray continuum, since their broad energy coverage allows for a more reliable treatment of absorption, Compton scattering, and reflection, especially for heavily obscured AGNs.

As an additional independent check of the intrinsic $2 - 10\ {\rm keV}$ fluxes adopted in our sample, we compare our estimates with the {\em NuSTAR} spectral analysis of~\cite{Serafinelli_2024} for the sources common to the two samples. For the sources marked with ``E'' in Table~\ref{tab:swift_bat_sample}, we infer an intrinsic flux from the intrinsic $2 - 10\ {\rm keV}$ luminosity reported by~\cite{Serafinelli_2024} according to Equation~\eqref{eq:intr_flux_lum} (we ignore K-correction in this case, since we do not have infromation about $\Gamma_X$) using the luminosity distance corresponding to the redshift and cosmology adopted in that work. These independently inferred fluxes are shown in parentheses in Table~\ref{tab:swift_bat_sample}. We emphasize that these values are used as a cross-check rather than replacing our fiducial estimates, allowing us to assess the differences arising from independent spectral analyses.

\begin{itemize}
\item IC 4329A, NGC 5506, MCG 05-23-16, ESO 103-35, Mrk 509: For these overlapping sources, the two estimates agree within a factor of approximately $2 - 3$, despite being based on different observations and spectral treatments.
\item NGC 3281: The largest discrepancy occurs for NGC 3281, for which our fiducial value, $F_{X,\rm intr}^{2-10\ {\rm keV}} \simeq 5.2 \times 10^{-11}\ {\rm erg\ cm^{-2}\ s^{-1}}$, is approximately a
factor of six larger than the value $\simeq 9.0 \times 10^{-12}\ {\rm erg\ cm^{-2}\ s^{-1}}$ inferred from the {\em NuSTAR} luminosity of~\cite{Serafinelli_2024}. NGC 3281 is a heavily obscured Seyfert galaxy, for which the authors employ an absorbed Comptonization continuum together with a torus-reflection model (their Model C), obtaining an intrinsic $2 - 10\ {\rm keV}$ luminosity of $(2.3 \pm 0.4) \times 10^{42}\ {\rm erg\ s^{-1}}$. The substantial difference between the two estimates highlights the model dependence of reconstructing the intrinsic X-ray emission in heavily obscured AGNs. We therefore retain our fiducial values while using the {\em NuSTAR} estimates to illustrate the possible source-dependent systematic differences.
\item GRS 1734-292: We compare our adopted flux with the {\em NuSTAR} analysis of~\citet{Tortosa:2016loq}. The latter reports an intrinsic $2-10\ {\rm keV}$ luminosity of $6.67 \times 10^{43}\ {\rm erg\ s^{-1}}$ for their 2014 {\em NuSTAR} observation, corresponding to an intrinsic flux of $\simeq 7.36 \times 10^{-11}\ {\rm erg\ cm^{-2}\ s^{-1}}$ using their quoted distance of $\simeq 87$ Mpc. This is approximately a factor of $1.5$ below our fiducial value of $1.095 \times 10^{-10}\ {\rm erg\ cm^{-2}\ s^{-1}}$. Their spectral analysis yields a photon index $\Gamma_X = 1.65 \pm 0.05$ and a high-energy cutoff
$E_{\rm cut} = 53^{+11}_{-8}\ {\rm keV}$, compared with our adopted {\em Swift}/BAT photon index of $\Gamma_X=2.21$ with an unbroken power-law. The difference may reflect variability between observing epochs and differences in spectral modeling. Since GRS 1734-292 is Compton-thin
($\log_{10}N_{\rm H} = 21.92$), uncertainties associated with correcting for Compton-thick absorption are not expected to dominate the inferred intrinsic continuum. We therefore retain the {\em Swift}/BAT-based flux and photon index for consistency with our sample methodology, while reporting the independent {\em NuSTAR} estimate in Table~\ref{tab:swift_bat_sample}.
\item 4U 1344-60: We compare our adopted flux with the independent {\em NuSTAR}-based analysis of~\cite{2025ApJ...994..151G}. The latter report an intrinsic $2 - 10\ {\rm keV}$ coronal luminosity of $ \simeq 1.7\times 10^{43}\ {\rm erg\ s^{-1}}$, corresponding to a flux of $\simeq 4.7 \times 10^{-11}\ {\rm erg\ cm^{-2} s^{-1}}$ using $\Gamma_X = 1.89$ and the luminosity distance of $\simeq 55$ Mpc inferred from their adopted redshift and cosmology. This is approximately a factor of $1.4$ below our fiducial intrinsic $2 - 10\ {\rm keV}$ flux of $6.64 \times 10^{-11}\ {\rm erg\ cm^{-2} s^{-1}}$. Unlike NGC 3281, 4U 1344-60 is Compton-thin ($\log_{10} N_H = 22.1$), and the difference may reflect variations between observing epochs and spectral modeling. We therefore retain the BASS/{\em Swift} BAT-based flux and photon index ($\Gamma_X=1.99$) for this source, consistent with our sample methodology.
\end{itemize}
\subsection{Details for particular candidates}
\label{subsec:details_cand}
In this subsection, we describe source-specific adjustments made during the construction of our sample.
\subsubsection{Source-specific adjustments for sources in Table~\ref{tab:swift_bat_sample}}
\label{subsubsec:source_specific}
\begin{itemize}
\item NGC 1068: The intrinsic $2 - 10$ keV luminosity of NGC 1068 is uncertain due to its Compton-thick nucleus. From the temporary unveiling of the nucleus observed with {\em NuSTAR},~\cite{Marinucci:2015fqo} inferred $\log_{10} (N_H/{\rm cm}^{-2}) \simeq 25$, $L_X^{2-10\ {\rm keV}} = 7^{+7}_{-4} \times 10^{43}\ {\rm erg\ s^{-1}}$, assuming $d_L = 14.4\ {\rm Mpc}$. Rescaling this luminosity to the recent Cepheid-based distance of $d_L = 10.7\ {\rm Mpc}$~\citep{2026ApJ..1000...78M} gives $L_X^{2-10\ {\rm keV}} \simeq 3.86 \times 10^{43}\ {\rm erg\ s^{-1}}$. This value is also consistent and lies within the uncertainties of subsequent {\rm NuSTAR} observations\footnote{A more recent analysis of broadband hard X-ray observations by~\cite{2026A&A...710A.144R}, employing different assumptions about the geometry of the obscuring material, suggests a substantially lower intrinsic luminosity. Understanding the obscuration and time variability is important.} from~\cite{Zaino:2020elj} who obtain $L_X^{2-10\ {\rm keV}} \simeq 1.9^{+2.0}_{-1.0} \times 10^{43}\ {\rm erg\ s^{-1}}$, after rescaling to the same distance ($d_L = 10.7$ Mpc). Therefore, we adopt the corresponding intrinsic flux $F_{X,\rm intr}^{2-10\ {\rm keV}} = 2.82 \times 10^{-9}\ {\rm erg\ cm^{-2} s^{-1}}$, corresponding to the rescaled $L_X^{2-10\ {\rm keV}}$ from~\cite{Marinucci:2015fqo}.

We choose $M_{\rm BH} = 8.0 \pm 0.3 \times 10^6 M_\odot$ from~\cite{Lodato:2002cv} where the SMBH mass was constrained using water maser kinematics assuming $d_L = 14.4\ {\rm Mpc}$. Using our simple scaling relation we recompute $M_{\rm BH}$ from our $d_L = 10.7\ {\rm Mpc}$~\citep{2026ApJ..1000...78M}. It is important to note that the $M_{\rm BH}$ that~\cite{Lodato:2002cv} is consistent with more recent measurements in~\cite{2020MNRAS.497.1020W} and~\cite{2024ApJ...975L...9G}.
\item Circinus Galaxy: We adopt $d_L = 4.2\ {\rm Mpc}$ and a fiducial value of $\Gamma_X = 2.3$ consistent with the broadband X-ray spectral analysis of the Compton-thick nucleus by~\cite{Arevalo:2014zva}. Their analysis inferred an intrinsic $2 - 10$ keV luminosity of $L_X^{2-10\ {\rm keV}} \simeq (2.3-5.1) \times 10^{42}\ {\rm erg\ s^{-1}}$, depending on the adopted spectral model. Thus, we adopt the fiducial value $L_X^{2-10\ {\rm keV}} = 10^{42.5}\ {\rm erg\ s^{-1}}$ as in~\cite{Murase:2023ccp}, which lies within this range. We note that the XCLUMPY torus analysis of~\cite{Tanimoto:2019tbz,Tanimoto:2023dzq} obtained a harder intrinsic spectrum, $\Gamma_X = 1.80$, and a somewhat higher luminosity, $L_X^{2-10\ {\rm keV}} \simeq 10^{42.8}\ {\rm erg\ s^{-1}}$, illustrating the model dependence of the inferred intrinsic emission. Our adopted luminosity corresponds to $F_{X,\rm intr}^{2-10\ {\rm keV}} \simeq 1.5 \times 10^{-9}\ {\rm erg\ cm^{-2}\ s^{-1}}$ at $d_L=4.2\ {\rm Mpc}$.

For the SMBH mass, we adopt the water-maser measurement $M_{\rm BH} = (1.7\pm0.3)\times10^6 M_\odot$ from~\cite{2003ApJ...590..162G}.

\item NGC 4945: The distance used for this source was quoted by~\citet{Tanimoto:2022wrq} to be $3.5$ Mpc, which is more accurate than the calculated distance using the redshift. On the other hand, their reported intrinsic X-ray luminosity is substantially lower than the result of another {\em NuSTAR} analysis by~\cite{2014ApJ...793...26P} (which is adopted by~\cite{Murase:2023ccp}). This could indicate significant variability in the intrinsic X-ray flux and column density associated with the source in addition to the adopted broadband spectral model.

We choose $M_{\rm BH}=1.4 \pm 0.7 \times 10^6\ M_\odot$ from~\cite{Greenhill:1996gz}, as in~\cite{Murase:2023ccp}, where the SMBH mass was determined from water-maser kinematics assuming $d_L=3.72~{\rm Mpc}$. Since the inferred dynamical mass scales linearly with the adopted distance, we rescale it to our adopted $d_L=3.55\ {\rm Mpc}$, obtaining $M_{\rm BH}=1.3 \pm 0.7 \times 10^6\ M_\odot$. 

\item NGC 4151: We adopt $\Gamma_X = 1.85$ and the intrinsic $2 - 10$ keV flux, $F_{X,\rm intr}^{2-10\ {\rm keV}} = 1.66 \times 10^{-10}\ {\rm erg\ cm^{-2}\ s^{-1}}$, from the simultaneous XMM-Newton, {\em NuSTAR}, and IXPE spectral analysis of~\cite{Gianolli:2023zji}. At our adopted distance of $d_L = 14.4\ {\rm Mpc}$, this corresponds to $L_X^{2-10\ {\rm keV}} \simeq 4.1 \times 10^{42}\ {\rm erg\ s^{-1}}$, consistent with the fiducial $L_X = 5\times 10^{42}\ {\rm erg\ s^{-1}}$ adopted in~\cite{Murase:2026hrz}. We use the measured $2 – 10$ keV flux to normalize the X-ray spectrum and extrapolate it to $2 – 195$ keV. For the SMBH mass, we adopt the velocity-resolved reverberation-mapping measurement $M_{\rm BH} = 1.66^{+0.48}{-0.34} \times 10^7 M\odot$ from~\cite{2022ApJ...934..168B}, which is also consistent with the fiducial $M_{\rm BH} = 1.7\times10^7 M_\odot$ considered in~\cite{Murase:2026hrz}.

We retain the intrinsic $2$–$10$ keV flux obtained from the {\em NuSTAR} spectral analysis of \cite{Shuvo:2022cxw}, since the measured flux is independent of the adopted source distance, the corresponding X-ray luminosity is instead calculated using our adopted $d_L$. 

\item NGC 7172: We adopt $M_{\rm BH} = (1.03 \pm 0.35) \times 10^7 M_\odot$, $\log_{10} (N_H/{\rm cm}^{-2}) = 24.14$, and $d_L = 37.5$ Mpc. Using the {\em Swift}/BAT $14 - 195$ keV flux with $\Gamma_X \simeq 1.9$, we infer intrinsic $2 - 10$ keV and $2 - 195$ keV fluxes of $76.7$ and $268$, respectively, in units of $10^{-12}\ {\rm erg\ cm^{-2}\ s^{-1}}$. These are consistent with the corresponding fluxes of $83.8$ and $279$ in the same units obtained using the {\em NuSTAR} spectral analysis of \cite{Shuvo:2022cxw}, which also gives $\Gamma_X \simeq 1.9$. We therefore adopt the {\em NuSTAR}-based fluxes in Table~\ref{tab:swift_bat_sample}.

\item MR 2251$-$178: We adopt the black-hole mass from~\cite{Lira_2011},
who estimated $M_{\rm BH} \simeq 200 \times 10^6\ M_\odot$ using the broad-line radius-luminosity relation together with the observed H$\beta$ line width, assuming $d_L=274\ {\rm Mpc}$. Since this single-epoch virial estimator scales linearly with the assumed $d_L$, we recompute the mass using our adopted $d_L=292\ {\rm Mpc}$, obtaining $M_{\rm BH}=213 \pm 53 \times 10^6\ M_\odot$.

\item NGC 6300: We adopt the black-hole mass estimate from~\cite{2012AstL...38..475K}, who inferred $M_{\rm BH}=39 \times 10^6\ M_\odot$ from the host-galaxy bulge luminosity assuming
$d_L=14.3\ {\rm Mpc}$. Unlike direct dynamical measurements, the bulge-luminosity method gives $M_{\rm BH}\propto L_{\rm bulge}^{0.925}$, where $L_{\rm bulge}\propto d_L^2$. Therefore, $M_{\rm BH}\propto d_L^{1.85}$.
Using our adopted $d_L=16.1\ {\rm Mpc}$, we therefore obtain $M_{\rm BH}=48 \times 10^6\ M_\odot$.

\item MCG-6-30-15: We adopt the stellar-dynamical measurement of $M_{\rm BH}=44 \pm 14 \times 10^6\ M_\odot$ from~\cite{das2025stellardynamicalmasscentral}, which assumes
$d_L=25.5~{\rm Mpc}$. Since stellar-dynamical $M_{\rm BH}$ scale linearly with the adopted distance, we recompute the mass using our adopted $d_L=33.7\ {\rm Mpc}$, obtaining $M_{\rm BH}=58 \pm 18 \times 10^6\ M_\odot$.
\end{itemize}
\subsubsection{Sources excluded from the final sample to compile Table~\ref{tab:swift_bat_sample}}
After placing a cut on the flux, we initially had 39 sources with flux higher than the cutoff. We individually checked each source and some of them were removed for the reasons mentioned below. After the changes were made, we had the 28 sources that are mentioned in Table~\ref{tab:swift_bat_sample}.
\begin{itemize}
\item We removed Cen A~\citep{marin2025scatteredsynchrotronemissiongiant}, Cygnus A~\citep{Reynolds_2015}, 3C 111~\citep{chernyakova2004integralrxteobservationsbroadline}, 3C 120~\citep{Chatterjee_2009}, and 3C 390.3~\citep{leighly1995resultsmonitoringbroadlineradio}, which are radio galaxies with prominent relativistic jets and therefore fall outside the non-jetted Seyfert-like AGN population considered here. 
\item We also excluded NGC 6240~\citep{Gerssen_2004}, which is a merging ultraluminous infrared galaxy hosting two active nuclei and therefore does not conform to the single-nucleus AGN framework adopted in this work. In addition, there is also star-forming activity associated with mergers~\citep{2010A&A...524A..56E,Nardini:2017qib}.
\item NGC 1275 and Mrk 348 were flagged as BZU and BZB, respectively, in the BASS sample and were therefore initially excluded. However, the 157-month {\em Swift}/BAT sample classifies both sources as Seyfert 1.9 galaxies, and they were consequently reintroduced with sample ID 2; both satisfied our flux criterion. We subsequently excluded NGC 1275 because it is a radio galaxy with prominent jet-associated nonthermal emission~\citep{Aleksi__2014}, placing it outside the non-jetted Seyfert-like population considered here. Mrk 348 was retained in the final sample.
\item UGC 10683 NED02 initially satisfied our flux criterion. But for this source the {\em Swift}/BAT 157-month catalog~\citep{Lien:2025zzo} reports a $\Gamma_X = 4.99$. Such an unusually steep $14 - 195$ keV X-ray spectrum, together with the substantial discrepancy between the catalog measurements, raises concerns about the reliability of the inferred spectral parameters. We therefore conservatively exclude this source from our final sample.

\item 2MASX J15120505$-$1046356 also initially satisfied our flux criterion. However this source is associated with a {\em Swift}/BAT detection region containing multiple AGN counterparts, including NVSS J151148$-$105023, which is classified as a blazar~\citep{2013ApJS..207...19B,Koss:2022lkb}. This introduces additional uncertainty in attributing the measured hard X-ray emission to the Seyfert nucleus. Given these uncertainties, we conservatively exclude this source from our final sample.
\item ESO 138-1 source had sample ID 1234, hence the flux was calculated using values from ~\citet{Tanimoto:2022wrq}. The calculated flux did not make the cut (see Equation~\ref{eq:intr_xray_flux_cut}).

\item NGC 4388: Using the {\em Swift}/BAT $14 - 195$ keV flux and the BASS photon index $\Gamma_X = 1.78$, we infer intrinsic $2 - 10$ keV and $2 - 195$ keV fluxes of $84.4$ and $346$, respectively, in units of $10^{-12}\ {\rm erg\ cm^{-2} s^{-1}}$. However, using the intrinsic $2 - 10$ keV flux and $\Gamma_X = 1.66$ from the {\em NuSTAR} spectral analysis of~\cite{Shuvo:2022cxw}, we obtain substantially lower corresponding fluxes of $14$ and $71.5$ in the same units. Since the latter $2 - 195$ keV flux falls below our adopted selection threshold, we exclude NGC 4388 from our final sample.
\end{itemize}
\subsection{Tiers and rankings}
\label{subsec:rank_tiers}
We rank the 28 AGNs in our X-HiNAC sample according to their intrinsic X-ray fluxes in the $2-10$ and $2-195$ keV bands, as summarized in Table~\ref{tab:swift_bat_sample}. The rankings in the two energy bands are not identical, highlighting the importance of broadband X-ray spectral characterization when identifying promising neutrino source candidates. However, these rankings should not be interpreted as a definitive ordering of sources. Intrinsic X-ray variability, uncertainties in spectral modeling and absorption corrections, and differences between observations obtained at different epochs can substantially affect $F_{X,\rm intr}$. This is particularly relevant for lower-ranked sources, whose intrinsic X-ray fluxes often differ by only a few percent, making their relative rankings sensitive to updated measurements.

To provide a more robust classification, we additionally divide the sources into four flux-based tiers (Tier I, II, III, and IV) independently for the $2-10$ and $2-195$ keV bands. For the $2 - 10$ keV band, Tiers I, II, III, and IV correspond to $F_{X,\rm intr} > 1000$, $150 < F_{X,\rm intr} \leq 1000$, $80 \leq F_{X,\rm intr} \leq 150$, and $F_{X,\rm intr} < 80$, respectively, where the fluxes are expressed in units of $10^{-12}\ {\rm erg\ cm^{-2}\ s^{-1}}$. For the $2 - 195$ keV band, the corresponding boundaries are $F_{X,\rm intr} > 1000$, $400 < F_{X,\rm intr} \leq 1000$, $200 \leq F_{X,\rm intr} \leq 400$, and $F_{X,\rm intr} < 200$. These thresholds are chosen to group sources with broadly comparable intrinsic X-ray fluxes rather than to represent distinct physical regimes.

We find that Tier I contains two sources in both energy bands, while Tier II has 4 (6), Tier III has 9 (8), and Tier IV has 13 (12) sources for the $2-10$ keV ($14-195$ keV) band. This classification groups sources with broadly comparable $F_{X,\rm intr}$, reducing the emphasis on their exact rankings while preserving the distinction between the brightest and fainter candidates. As evident from Table~\ref{tab:swift_bat_sample}, some sources occupy different tiers depending on the adopted energy band. For example, NGC 4945 belongs to Tier III in the $2 - 10$ keV band but Tier II in the $2 - 195$ keV band. Nevertheless, 23 of the 28 sources ($\sim 80\%$) retain the same tier classification in both bands, indicating that the broader flux-based grouping is relatively insensitive to the choice of energy band.

We emphasize that even the highest tier (Tier I) sources are subject to substantial uncertainties. In particular, we explicitly discuss in Section~\ref{subsubsec:source_specific} that the intrinsic X-ray luminosities of the Compton-thick nuclei NGC 1068 and Circinus, both classified as Tier I, depend sensitively on the adopted spectral models and absorption corrections. Consequently, the tier assignments should be regarded as indicative rather than definitive, and the associated uncertainties must be considered when interpreting the predicted neutrino fluxes and detection prospects.
\section{Model-dependent prospects}
\label{sec:res}
In this section, we present the results for the neutrino spectra, event distribution, and $p$-value for both a sample-based search and a stacked search.
\subsection{Neutrino spectra}
\begin{figure}
\centering
\includegraphics[width=0.98\linewidth]{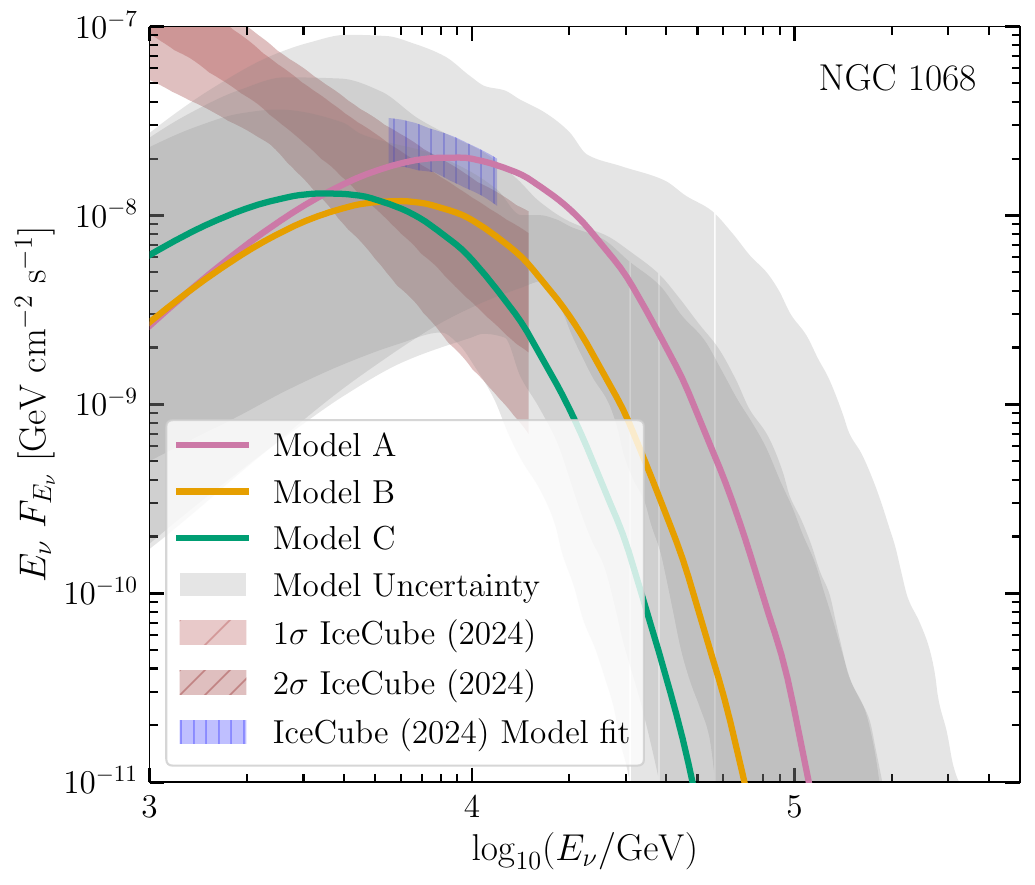}
\caption{\label{fig:ngc1068_spectra}Predicted all-flavor neutrino fluxes from NGC 1068 for Models A, B, and C, compared with the IceCube observations. The gray shaded regions indicate the model uncertainties obtained by propagating the uncertainty in the intrinsic X-ray luminosity~\citep{Murase:2026hrz}. The dark and light brown hatched regions represent the $1\sigma$ and $2\sigma$ regions, respectively, for the power-law neutrino flux inferred by IceCube~\citep{IceCube:2024ayt}, while the blue hatched region shows the corresponding IceCube disk-corona model-fit uncertainty.
}
\end{figure}
\begin{figure*}
\includegraphics[width=0.95\linewidth]{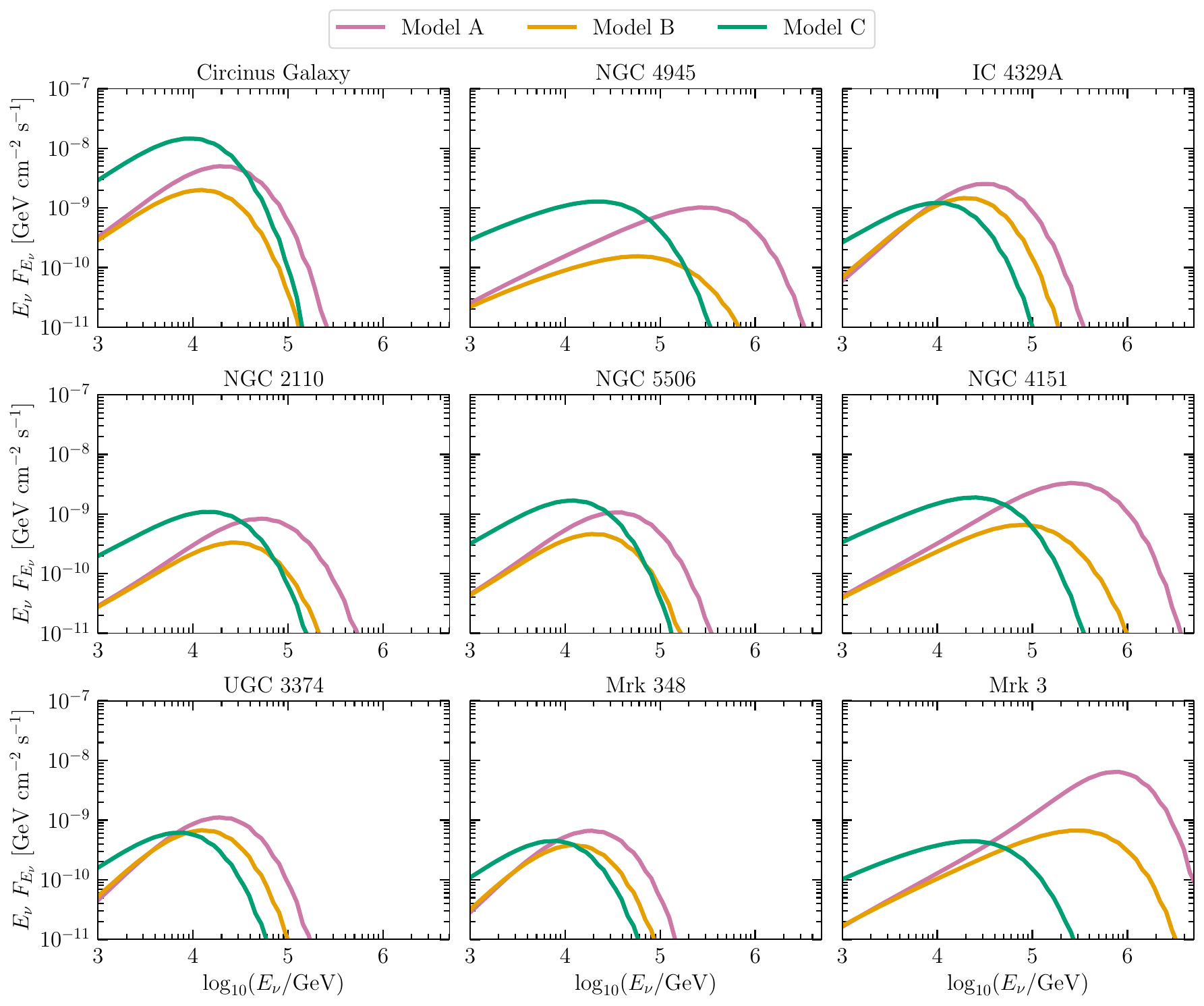}
\caption{\label{fig:nu_spectra}Predicted all-flavor neutrino energy fluxes for nine selected X-HiNAC sources (Table~\ref{tab:swift_bat_sample}), comprising the four highest-ranked sources in the Southern and Northern sky and one in the Horizon region. The purple, orange, and green curves correspond to Models A, B, and C, respectively. The spectra are calculated using the magnetically powered corona model, adopting a dimensionless coronal radius $\mathcal{R} = 3$ for $\lambda_{\rm Edd} \geq 0.3$ and $\mathcal{R}=10$ for $\lambda_{\rm Edd}<0.3$ for Models A and B, while for Model C we fix $\mathcal{R} = 10$ for all values of $\lambda_{\rm Edd}$.
}
\end{figure*}
We calculate the all-flavor neutrino spectra for the X-HiNAC sample in Table~\ref{tab:swift_bat_sample} using the magnetically powered corona model described in Section~\ref{sec:nu_agn}, following the treatment of~\cite{Murase:2026hrz}. For each source, we convert the intrinsic $2 - 10\ {\rm keV}$ X-ray flux into luminosity using $L_X = 4\pi d_L^2 F_{X,\rm intr}^{2-10\,{\rm keV}}$, and adopt the source-specific X-ray spectral parameters and black hole masses compiled in our sample.

Given the significant detection of neutrino emission from NGC 1068 by IceCube~\citep{IceCube:2022der}, we separately compare its predicted neutrino spectra with the observational results in Figure~\ref{fig:ngc1068_spectra}. As discussed in Section~\ref{subsec:details_cand}, the intrinsic X-ray luminosity of NGC 1068 is uncertain due to its heavily obscured nucleus. We propagate this uncertainty into the predicted neutrino spectra for each model, shown by the gray shaded envelopes. For comparison, we also show the $1\sigma$ and $2\sigma$ regions of the power-law fit inferred by IceCube, together with the model-dependent fit~\citep{IceCube:2024ayt}. The latter follows a template flux for a specific set of parameters ($L_X = 10^{42.9}\ {\rm erg\ s^{-1}}$ from~\citealt{Koss:2017ilk} and the ratio of CR pressure to thermal pressure $P_{\rm CR}/P_{\rm th} = 0.5$) for the magnetized corona model and should therefore be distinguished from Models A, B, and C considered in this work. We find that the predicted spectra are broadly compatible with the IceCube results within the uncertainties, although the agreement depends on the assumed model and energy range. In particular, the different spectral shapes and high-energy cutoffs illustrate the limitations of a simple power-law description and motivate future model-dependent analyses of the neutrino emission from NGC 1068.

The predicted all-flavor neutrino energy fluxes for nine representative AGNs, including the four highest-ranked sources in the northern and southern sky and one in the horizon, based on their intrinsic $2 - 195\ {\rm keV}$ X-ray fluxes (Table~\ref{tab:swift_bat_sample}) are shown in Figure~\ref{fig:nu_spectra}. The predicted spectra exhibit substantial variations in both normalization and spectral shape across the three models and individual sources. In general, Model C produces relatively stronger emission at lower neutrino energies ($E_\nu \sim 1 - 30\ {\rm TeV}$), but exhibits a lower high-energy cutoff. Model A typically extends to higher energies, reaching several hundred TeV or beyond for some sources, while Model B generally predicts lower fluxes. Among the selected sources, Circinus Galaxy and NGC 1068 exhibit particularly large neutrino fluxes, although their relative strengths depend on the assumed model. These differences highlight the importance of the underlying coronal properties in determining the predicted neutrino emission. Since the detector effective areas depend strongly on neutrino energy and source declination, the relative detectability of these sources cannot be inferred from their flux normalizations alone, motivating the event rate calculations presented in the following subsection.

The neutrino emission from Circinus Galaxy, NGC 4945, and NGC 4151 has also been investigated in previous multimessenger studies~\citep{Murase:2023ccp,Carpio:2026xkf}. In particular, \cite{Murase:2023ccp} considered two representative coronal configurations for these sources, the first adopting one of the parameter sets previously used to predict neutrino emission from NGC 1068~\citep{Murase:2019vdl,IceCube:2022der}, and the second assuming a more compact emission region with an enhanced CR pressure. For Circinus Galaxy and NGC 4945, the latter configuration predicts stronger neutrino emission. More recently,~\cite{Carpio:2026xkf} performed a Markov Chain Monte Carlo (MCMC) analysis of selected AGNs, jointly constraining the coronal parameters using neutrino and gamma-ray observations. The corresponding neutrino spectral bands were obtained by sampling parameter combinations from the MCMC posterior, thereby accounting for variations in the coronal properties. We find that our predicted neutrino spectra for Circinus Galaxy lie within the $1\sigma$ spectral bands reported in that work, despite adopting fixed coronal parameters and independently compiled X-ray properties.

For NGC 4151, we note that Model A predicts a relatively hard neutrino spectrum peaking at $E_\nu \sim 200 - 300$ TeV. Such a spectrum may not adequately reproduce the lower-energy neutrino excess inferred from the power-law fit reported by IceCube, as also discussed by~\cite{Carpio:2026xkf}. However, this does not necessarily exclude Model A, since the predicted number of events at PeV energies remains small, and the absence of detected PeV neutrinos does not by itself provide a significant constraint of this scenario. This illustrates the importance of accounting for the energy-dependent detector response when assessing the compatibility of physically motivated neutrino spectra with current observations.

\subsection{Event distributions}
\label{subsec:ev_dist}
\begin{figure*}
\includegraphics[width=0.98\linewidth]{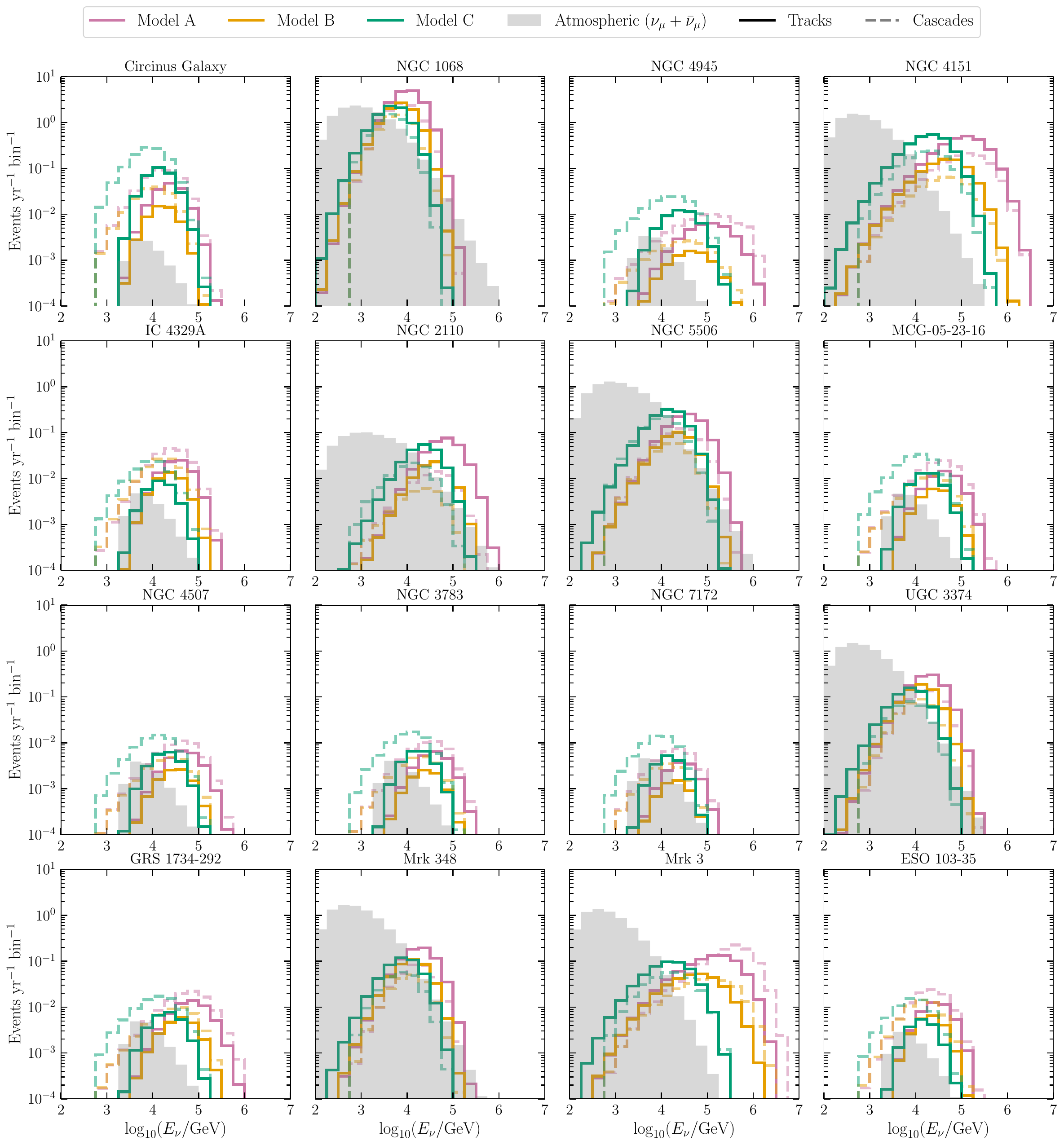}
\caption{\label{fig:events_icg2_1}Predicted muon-neutrino ($\nu_\mu+\bar{\nu}_\mu$) event rates for the first 16 AGNs in Table~\ref{tab:swift_bat_sample}
at \textbf{IceCube-Gen2}. The purple, orange, and green histograms correspond to Models A, B, and C, respectively, while the gray shaded histograms show the expected atmospheric-neutrino background. The event rates are shown per year, using four logarithmic energy bins per decade. The dashed lines show the event rates for the cascade channel. The signal and background are calculated using the IceCube-Gen2 effective area, with the background integrated over the energy-dependent angular resolution of the detector (see Section~\ref{subsec:ev_dist}). Also note that the event rates are shown as functions of the true neutrino energy ($E_\nu$), rather than the reconstructed muon energy.
}
\end{figure*}
\begin{figure*}
\includegraphics[width=0.98\linewidth]{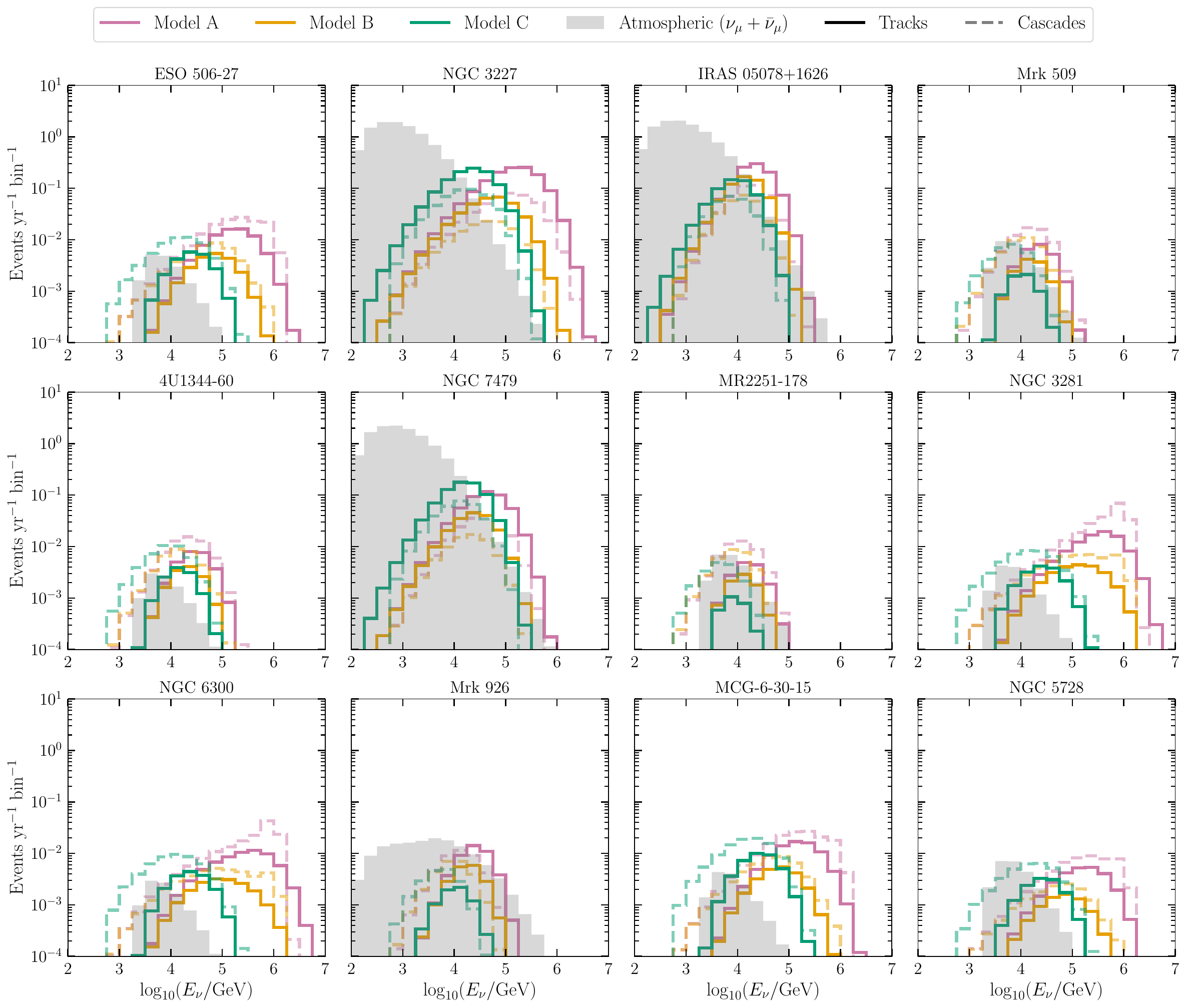}
\caption{\label{fig:events_icg2_2}Same as Figure~\ref{fig:events_icg2_1} for \textbf{IceCube-Gen2}, but for the remaining AGNs in Table~\ref{tab:swift_bat_sample}.
}
\end{figure*}
\begin{figure*}
\includegraphics[width=0.98\linewidth]{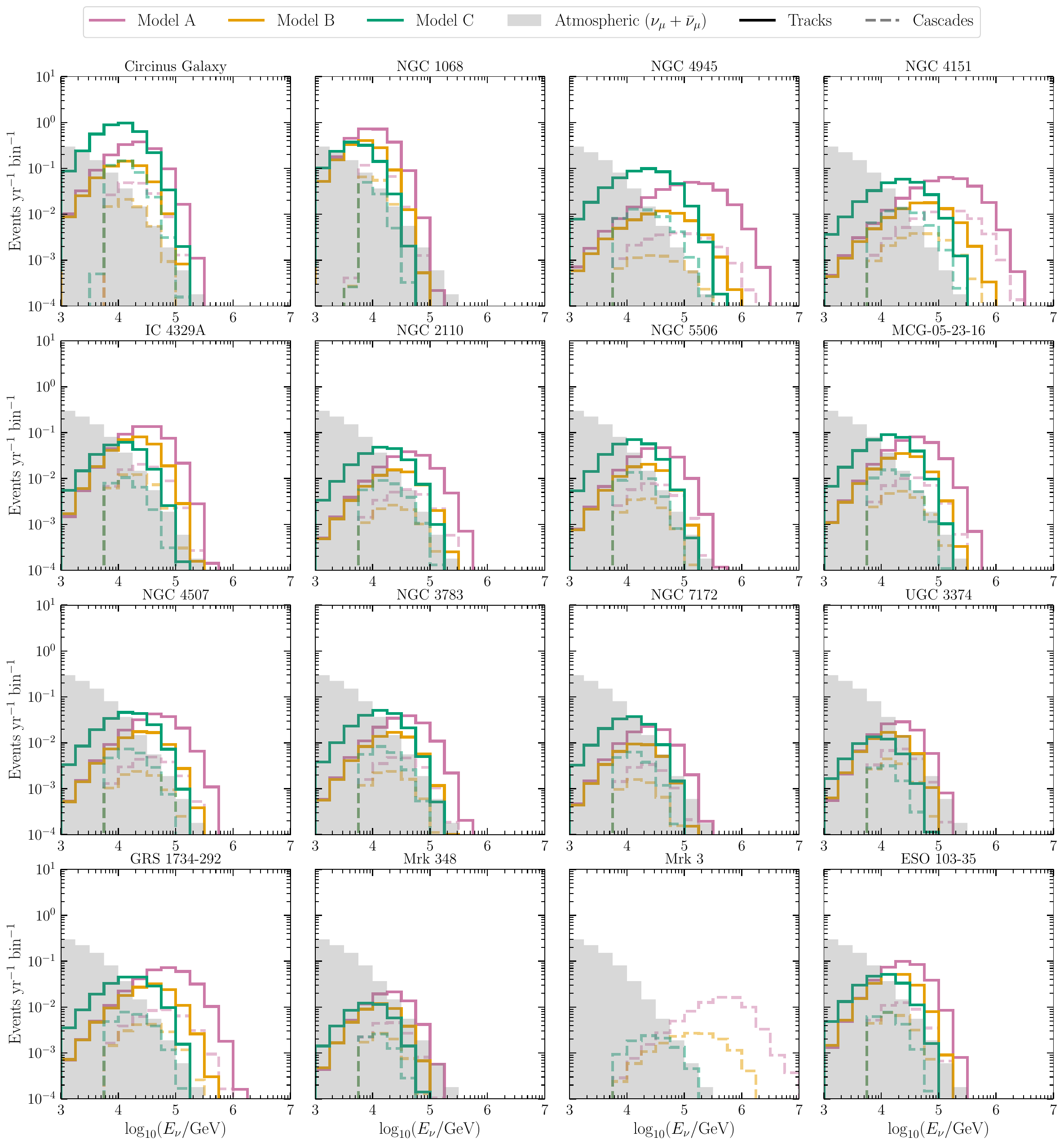}
\caption{\label{fig:events_km3net_1}Same as Figure~\ref{fig:events_icg2_1}, but for \textbf{KM3NeT}. The event rates are calculated using the KM3NeT effective area and the source-dependent visibility factor, accounting for the fraction of time each source is below the horizon (see Section~\ref{subsec:ev_dist}).
}
\end{figure*}
\begin{figure*}
\includegraphics[width=0.98\linewidth]{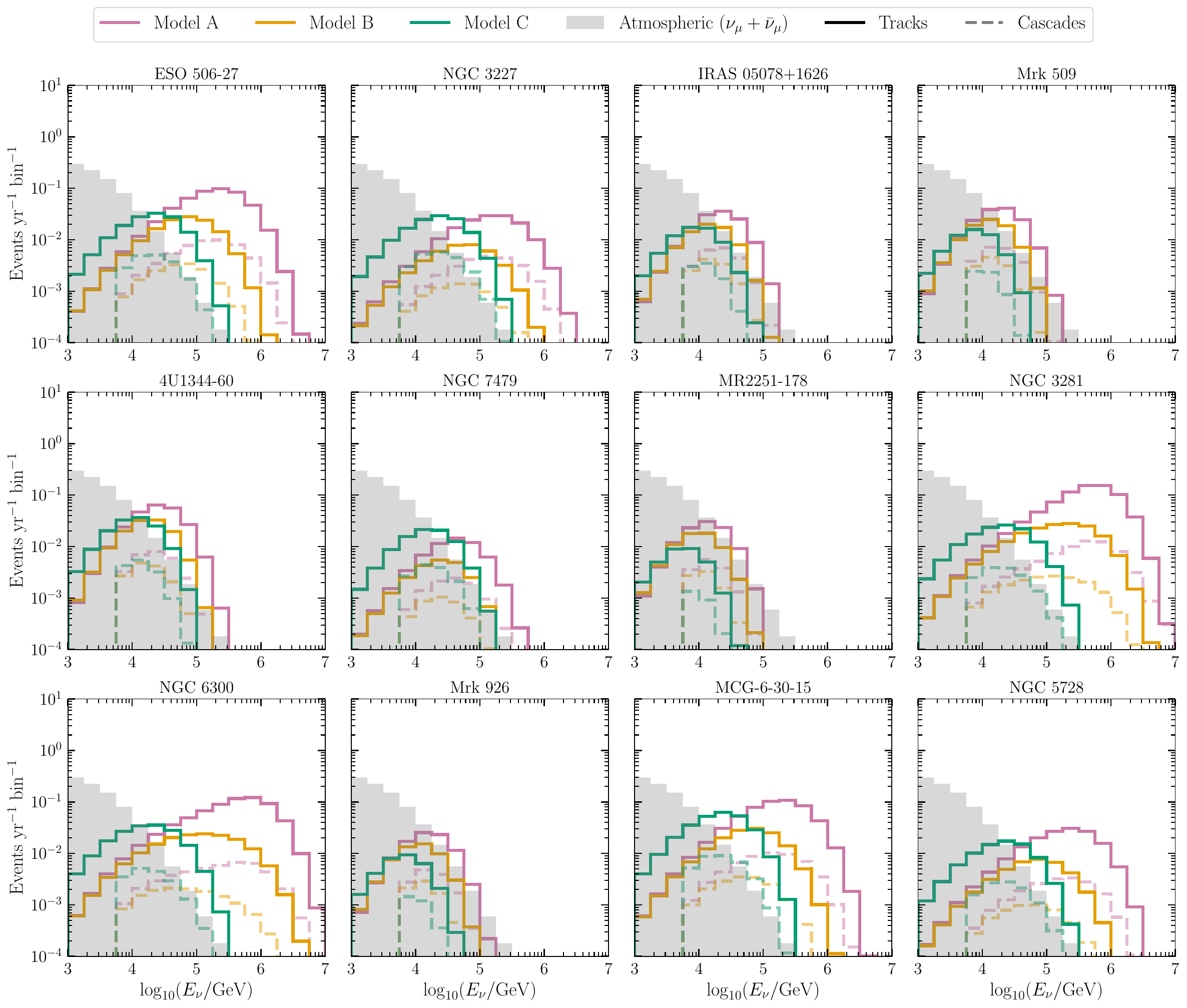}
\caption{\label{fig:events_km3net_2}Same as Figure~\ref{fig:events_km3net_1} for \textbf{KM3NeT}, but for the remaining AGNs in Table~\ref{tab:swift_bat_sample}.
}
\end{figure*}
In this subsection, we discuss the signal and background event distribution for the upcoming neutrino telescopes IceCube-Gen2 and KM3NeT. For signal events, we fold the neutrino fluxes with the effective areas $A_{\rm eff}$ from the corresponding neutrino detector, for a given period of operation $T_{\rm op}$. 

\begin{equation}
N = T_{\rm op}\int dE\ A_{\rm eff}(E)\Phi(E).
\end{equation}

We choose $T_{\rm op} = 10$ years for IceCube-Gen2 and KM3NeT. We consider the $\nu_\mu+\bar\nu_\mu$ effective areas, given the higher angular resolution of tracks compared to cascades. For IceCube, this will correspond to the PSTracks effective area from the IceCube data release~\citep{IceCube:2021xar} for source declinations $\delta>-10^\circ$, and the ESTES effective area when $\delta<-10^\circ$ \citep{IceCube:2025zyb}, while for KM3NeT the effective area is from~\cite{KM3NeT:2024paj}. In the case of IceCube-Gen2, we upscale the IceCube effective areas by a factor of $10^{2/3}$ to account for the increase in fiducial volume~\citep{IceCube-Gen2:2020qha,Gen2_TDR}.

For the signal flux, we use 1/3 of the all-flavor fluxes $\Phi$ shown in Figure~\ref{fig:nu_spectra}, assuming a $1:1:1$ $\nu+\bar\nu$ flavor ratio. The results for the signal events in each case are shown in Figures~\ref{fig:events_icg2_1},~\ref{fig:events_icg2_2} and~\ref{fig:events_km3net_1},~\ref{fig:events_km3net_2} for IceCube-Gen2 and KM3NeT respectively. We see that sources in the Southern sky (e.g., Circinus Galaxy, NGC 4945) have low event rates in IceCube due to the decreased $A_{\rm eff}$ at those declinations. 

The contribution from background events is also shown in gray in the same figure. We consider the background events to consist of atmospheric neutrinos, using the conventional atmospheric $\nu_\mu +\bar\nu_\mu$ flux from~\citet{Honda:2015fha}. We neglect the contribution of atmospheric muons, as the IceCube tracks dataset is most sensitive in the northern sky, where tracks are upgoing and atmospheric muons are attenuated. The IceCube ESTES dataset has negligible contributions from atmospheric muons, as the events consist of starting tracks. In KM3NeT, we can neglect atmospheric muons by only accounting for the times when the source would be located below the horizon. This fraction is accounted for as a visibility factor provided in \cite{KM3Net:2016zxf}. With IceCube located at the South Pole, we use the zenith-dependent Honda flux, while KM3NeT will use the sky-averaged Honda flux. 

Given the angular resolution of the detector $\sigma$ we integrate the atmospheric neutrino flux over a solid angle $\Omega_{\rm res} (\sigma) = \pi \sigma^2$. We adopt $\sigma=0.4^\circ$ for IceCube-Gen2 and KM3NeT, respectively. This assumes that IceCube and IceCube-Gen2 have similar angular resolutions. Similar to the signal events, the background neutrino spectrum is folded in with the effective area for each detector.

The predicted signal and atmospheric background event
distributions for the AGNs in our sample are shown in Figures~\ref{fig:events_icg2_1} - \ref{fig:events_km3net_2}. The event rates depend strongly on source declination and
detector location. For IceCube-Gen2, NGC 1068 and NGC 4151 are among the most promising sources, with the predicted event rates depending strongly on the assumed emission model. In contrast, the event rates from Southern-sky sources, such as Circinus Galaxy and NGC 4945, are suppressed by the smaller effective area at these declinations. For KM3NeT, the situation is reversed, with Circinus Galaxy emerging as one of the brightest sources, particularly for Model C. These differences highlight the complementary sky
coverage of the two observatories. Note that, Mrk 3 has no appreciable signal in KM3NeT. At a declination of $\delta=+73^\circ$, this source is not below KM3NeT's horizon, making its visibility drop to zero.

For reference, we also calculate the expected number of cascade events in IceCube-Gen2  and KM3NeT, from each of our sources. We use the DNNCascade $A_{\rm eff}$ as presented in~\cite{IceCube:2025lev} for IceCube events and the shower $A_{\rm eff}$ in~\cite{KM3NeT:2024paj} for KM3NeT. Similar to tracks, we fold the effective area with the per-flavor source flux, assuming a 1:1:1 flavor ratio. To account for the KM3NeT visibility factor, we calculate the number of cascade events above (below) the horizon, $N^{\rm casc}_{\cos\theta>0} (N^{\rm casc}_{\cos\theta<0})$, where $\theta$ is the zenith angle, using the corresponding effective areas. Therefore, if we denote the visibility factor as $\mu$, then $N_{\rm sig}^{\rm casc} = \mu N^{\rm casc}_{\cos\theta<0} + (1-\mu)N^{\rm casc}_{\cos\theta>0}$. These cascade event distributions are not used to calculate a $p$-value, as their angular uncertainty is significantly larger than track events. We show the signal event distribution in the cascade channel for IceCube-Gen2 (Figures~\ref{fig:events_icg2_1} and~\ref{fig:events_icg2_2}) and KM3NeT (Figures~\ref{fig:events_km3net_1} and~\ref{fig:events_km3net_2}) for each model using dashed lines.

The event distributions exhibit substantial model dependence. Model C generally produces higher event rates at lower energies, whereas Model A extends to higher energies and often dominates above $\sim 100\ {\rm TeV}$. Although the atmospheric neutrino background is generally larger at lower energies, it decreases rapidly with energy, making the higher energy events particularly important for identifying an astrophysical contribution.

In IceCube-Gen2, we typically get more cascade events than tracks for sources in the southern Sky, while the opposite is true for northern sky sources. This is explained by the effective area in ESTES and PSTracks being generally lower than in cascades for Southern sky sources.

We emphasize that the event distributions are calculated using the true neutrino energy, without accounting for energy reconstruction. For track events, the reconstructed muon energy is an imperfect proxy for the parent neutrino energy, particularly for through-going muons that lose energy before reaching the detector. Consequently, the pronounced high-energy peaks, especially near PeV energies, would be broadened and shifted toward lower reconstructed energies (see Figure 4 of~\citealt{Abbasi:2026ehs}). Thus, these distributions should not be interpreted as directly observable reconstructed energy spectra.
\subsection{Stacked and catalog-based searches}
\begin{table*}[ht]
\centering
\begin{tabular}{lcccccc}
\hline
& \multicolumn{3}{c}{IceCube-Gen2}
& \multicolumn{3}{c}{KM3NeT} \\
\cline{2-4} \cline{5-7}
Source name
& Model A & Model B & Model C
& Model A & Model B & Model C \\
\hline
\hline
NGC 1068
& $4.795\times10^{-23}$ & $8.00\times10^{-8}$ & $9.11\times10^{-7}$
& $1.13\times10^{-7}$ & $1.57\times10^{-3}$ & $3.63\times 10^{-3}$ \\
\hline
Circinus Galaxy
& $4.35\times10^{-2}$ & 0.50 & $1.83\times10^{-3}$
& $1.57\times10^{-3}$ & 0.107 & $9.05\times10^{-12}$ \\
& (0.50) & (0.50) & 
& (0.176) & (0.390) & \\

\hline
NGC 4945
& 0.50 & 0.50 & 0.50
& 0.271 & 0.473 & 0.176 \\
& (0.50) & (0.50) &  
& (0.390) & (0.473) &  \\
\hline
NGC 4151
& $1.88\times10^{-2}$ & 0.262 & $2.44\times10^{-2}$
& 0.271 & 0.390 & 0.271 \\
& (0.262) & (0.430) &  
& (0.390) & (0.473) &   \\
\hline
IC 4329A
& $7.76\times10^{-2}$ & 0.50 & 0.50
& 0.107 & 0.271 & 0.271 \\
\hline
NGC 2110
& 0.292 & 0.431 & 0.292
& 0.390 & 0.473 & 0.390 \\
& (0.431) & (0.431) &  
& (0.473) & (0.473) &   \\
\hline
NGC 5506
& 0.203 & 0.400 & 0.145
& 0.390 & 0.390 & 0.271 \\
& (0.400) &   & ($1.51\times10^{-2}$)
& (0.390) & (0.473) &   \\
\hline
MCG-05-23-16
& 0.50 & 0.50 & 0.50
& 0.271 & 0.390 & 0.176 \\
& (0.50) & (0.50) &  
& (0.390) & (0.473) &  \\
\hline
NGC 4507
& 0.50 & 0.50 & 0.50
& 0.390 & 0.390 & 0.390 \\
& (0.50) & (0.50) &  
& (0.390) & (0.473) &   \\
\hline
NGC 3783
& 0.50 & 0.50 & 0.50
& 0.390 & 0.473 & 0.390 \\
& (0.50) & (0.50) &  
& (0.473) & (0.473) &   \\
\hline
NGC 7172
& 0.50 & 0.50 & 0.50
& 0.390 & 0.473 & 0.390 \\
& (0.50) & (0.50) & 
& (0.473) & (0.473) & \\
\hline
UGC 3374
& 0.187 & 0.295 & 0.337
& 0.390 & 0.473 & 0.473 \\
\hline
GRS 1734-292
& 0.50 & 0.50 & 0.50
& 0.271 & 0.390 & 0.390 \\
& (0.50) & (0.50) & (0.50)
& (0.390) & (0.473) & (0.176) \\
\hline
Mrk 348
& 0.276 & 0.394 & 0.353
& 0.390 & 0.473 & 0.473 \\
\hline
Mrk 3
& 0.238 & 0.416 & 0.321
& 0.473 & 0.473 & 0.473 \\
& (0.416) & (0.466) & (0.139)
& (0.473) & (0.473) & (0.473) \\
\hline
ESO 103-35
& 0.50 & 0.50 & 0.50
& 0.176 & 0.390 & 0.473 \\
\hline
ESO 506-27
& $8.43\times10^{-2}$ & 0.50 & 0.50
& 0.107 & 0.390 & 0.390 \\
& (0.50) & (0.50) & (0.50)
& (0.390) & (0.473) & (0.176) \\
\hline
NGC 3227
& 0.167 & 0.428 & 0.220
& 0.390 & 0.473 & 0.390 \\
& (0.389) & (0.467) & ($1.75\times10^{-2}$)
& (0.473) & (0.473) & (0.176) \\
\hline
IRAS 05078+1626
& 0.234 & 0.330 & 0.365
& 0.390 & 0.390 & 0.473 \\
\hline
Mrk 509
& 0.50 & 0.50 & 0.50
& 0.390 & 0.390 & 0.473 \\
\hline
4U1344-60
& 0.50 & 0.50 & 0.50
& 0.271 & 0.390 & 0.390 \\
\hline
NGC 7479
& 0.374 & 0.445 & 0.308
& 0.473 & 0.473 & 0.390 \\
& (0.445) & (0.482) & (0.149)
& (0.473) & (0.482) & (0.390) \\
\hline
MR 2251-178
& 0.50 & 0.50 & 0.50
& 0.390 & 0.473 & 0.473 \\
\hline
NGC 3281
& $6.88\times10^{-2}$ & 0.50 & 0.50
& $6.10\times10^{-2}$ & 0.390 & 0.390 \\
& (0.50) & (0.50) & (0.50)
& (0.390) & (0.473) & (0.176) \\
\hline
NGC 6300
& 0.50 & 0.50 & 0.50
& $6.10\times10^{-2}$ & 0.390 & 0.390 \\
& (0.50) & (0.50) & (0.50)
& (0.390) & (0.473) & (0.107) \\
\hline
Mrk 926
& 0.342 & 0.338 & 0.338
& 0.390 & 0.473 & 0.473 \\
\hline
MCG-6-30-15
& $6.96\times10^{-2}$ & 0.50 & 0.50
& 0.107 & 0.390 & 0.271 \\
& (0.50) & (0.50) &  
& (0.390) & (0.473) &   \\
\hline
NGC 5728
& 0.50 & 0.50 & 0.50
& 0.390 & 0.473 & 0.390 \\
& (0.50) & (0.50) &  
& (0.473) & (0.473) &  \\
\hline
\hline
\end{tabular}
\caption{\label{tab:pvalues_gen2_km3net}$p$-values for track and cascade analyses for IceCube-Gen2 and KM3NeT. For Models A and B, we also show the $p$-value when $\mathcal{R} = 30$ for $\lambda_{\rm Edd} < 0.3$ in parentheses.
}
\end{table*}
\begin{figure*}
\centering
\includegraphics[width=0.95\linewidth]{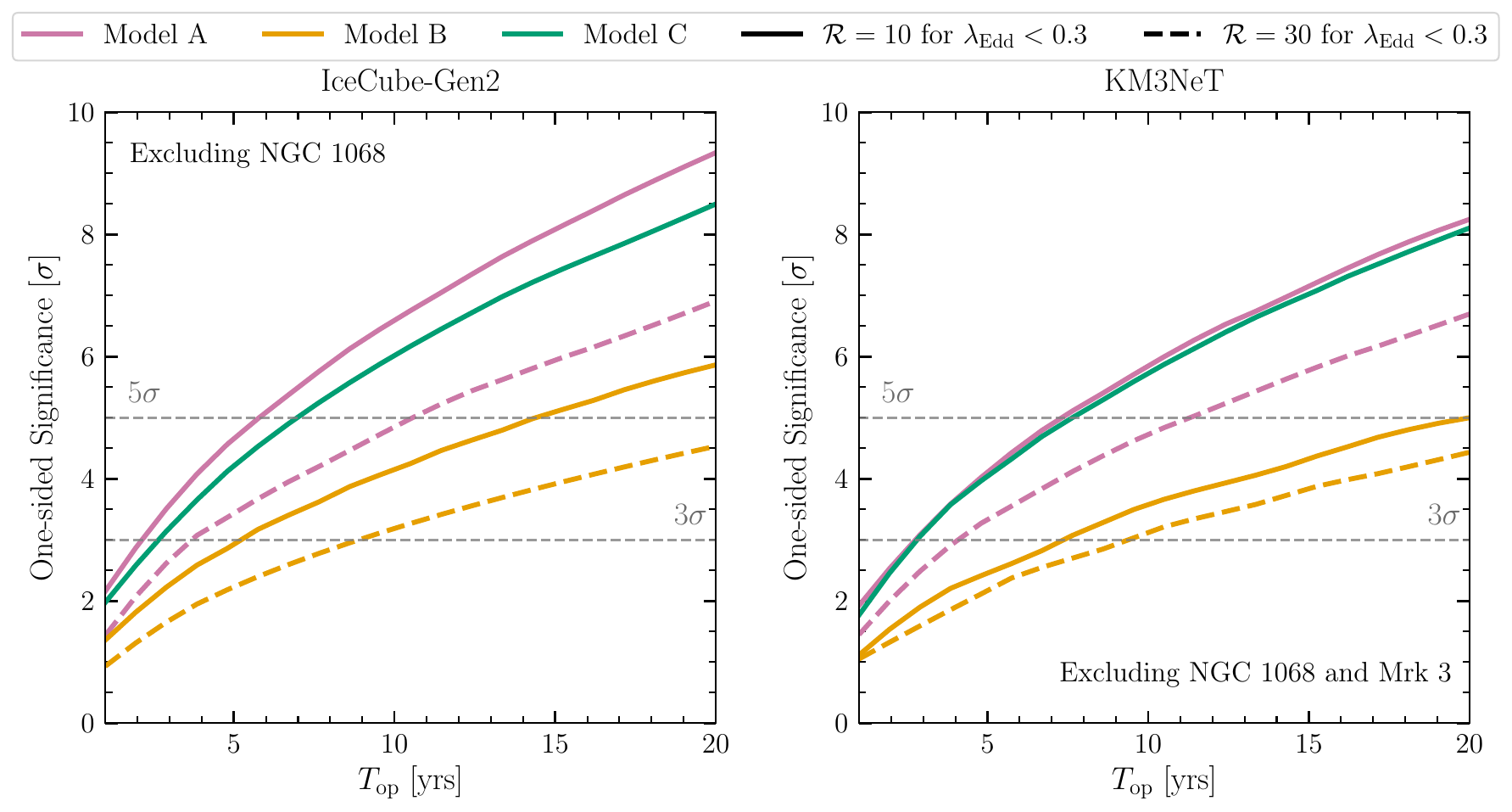}
\caption{\label{fig:sig}Projected one-sided statistical significance of the combined neutrino signal from the X-HiNAC sample in Table~\ref{tab:swift_bat_sample}, excluding NGC 1068, as a function of operation time for IceCube-Gen2 (left) and KM3NeT (right). For the sources where $\lambda_{\rm Edd} < 0.3$ we show the scenario of $\mathcal{R} = 10$ in solid and $\mathcal{R} = 30$ in dashed (Model C is only valid for $\mathcal{R} = 10$ for $\lambda_{\rm Edd}<0.3$).
}
\end{figure*}
In this section, as in~\citet{Kheirandish:2021wkm}, we investigate the prospects for detecting neutrino emission from the X-HiNAC sample in Table~\ref{tab:swift_bat_sample} through individual-source and stacking searches with IceCube-Gen2 and KM3NeT.

We estimate the statistical significance using a simplified Poisson counting analysis, following the prescription of~\citet{CMS-NOTE-2011-005}. For an individual source ($j$) the test statistic is defined as
\be
\label{eq:q0_individual}
q_{0,j} = \begin{cases}
2\Bigg( Y_{b,j}-N_{D,j} + N_{D,j} \ln \left(\dfrac{N_{D,j}}{Y_{b,j}} \right)
\Bigg), & \hspace{-1em} N_{D,j}>Y_{b,j},\\
0, & \hspace{-5em} N_{D,j}=0\; {\rm or }\; Y_{b,j}=0,
\end{cases}\,
\ee
where $Y_{b,j}$ is the expected number of atmospheric background events and $N_{D,j}$ is the median of the Poisson distribution with mean equal to the expected signal-plus-background event count for source $j$. We consider the total number of events between $100\ {\rm GeV}$ and $10\ {\rm PeV}$, treating the entire energy interval as a single bin. This approach avoids relying on reconstructed energy distributions and the associated energy smearing, at the expense of discarding potentially useful spectral information. The background is calculated within the angular search region described in the previous subsection, and we assume a signal containment fraction of $72\%$~\citep{Alexandreas:1992ek}. For KM3NeT, we note that the effective area reported in~\cite{KM3NeT:2024paj} is not provided below $1\ {\rm TeV}$.

For individual sources, we estimate the one-sided $p$-value using
\be
\label{eq:pvalue_individual}
p_j = \frac{1}{2} \operatorname{erfc} \left( \sqrt{\frac{q_{0,j}}{2}}\right)\,.
\ee
The corresponding projections for $10$ years of operation time ($T_{\rm op}$) are summarized in Table~\ref{tab:pvalues_gen2_km3net}. The detection prospects vary considerably among sources and emission models, reflecting differences in the predicted neutrino spectra, source declinations, and detector responses. For IceCube-Gen2, NGC 1068 is the most promising individual source, reaching a projected significance of approximately $7\sigma$ under Model A and $3.7\sigma$ under Model B. For KM3NeT, Circinus Galaxy is the most promising individual source under Model C, with a projected significance of approximately $6.7\sigma$ after 10 years, while NGC 1068 also exhibits promising detection prospects under the same model. Most other sources have considerably weaker individual detection prospects within the assumed observation period.

We further investigate the cumulative neutrino signal through
a stacking analysis. The test statistic in this case is defined by Equation~\eqref{eq:pvalue_individual}, but replacing $Y_{b,j}\to\sum_j Y_{b,j}$ and $N_{D,j}$ with the Poisson median with Poisson mean equal to the total signal-plus-background events from all sources. In Figure~\ref{fig:sig} we present the resulting one-sided significance as a function of $T_{\rm op}$ for IceCube-Gen2 and KM3NeT, excluding NGC 1068. We exclude this source to assess the detection prospects for the remaining AGNs independently of one of the most prominent individual candidates. While most sources yield modest individual significances, their combined contribution can be more readily detectable. For both detectors, Models A and C reach a nominal significance of approximately $5\sigma$ within $\sim 5 - 10$ years, whereas Model B may need a substantially longer exposure. Remarkably, these results suggest that a collective neutrino signal from the sample could be detectable even without NGC 1068.

We emphasize that these estimates are based on an approximate, energy-integrated counting analysis rather than a full event-level likelihood treatment. In particular, they do not incorporate the reconstructed energy and directional probability distributions, or systematic uncertainties in the detector response and atmospheric background. Furthermore, the nominal stacked significances obtained require an appropriate calibration of the combined test statistic under the background-only hypothesis. Our results should therefore be interpreted as indicative forecasts motivating dedicated catalog-based and stacking searches with next-generation neutrino telescopes.

\section{Conclusion and outlook}
\label{sec:concl}
\begin{figure*}
\includegraphics[width=0.95\linewidth]{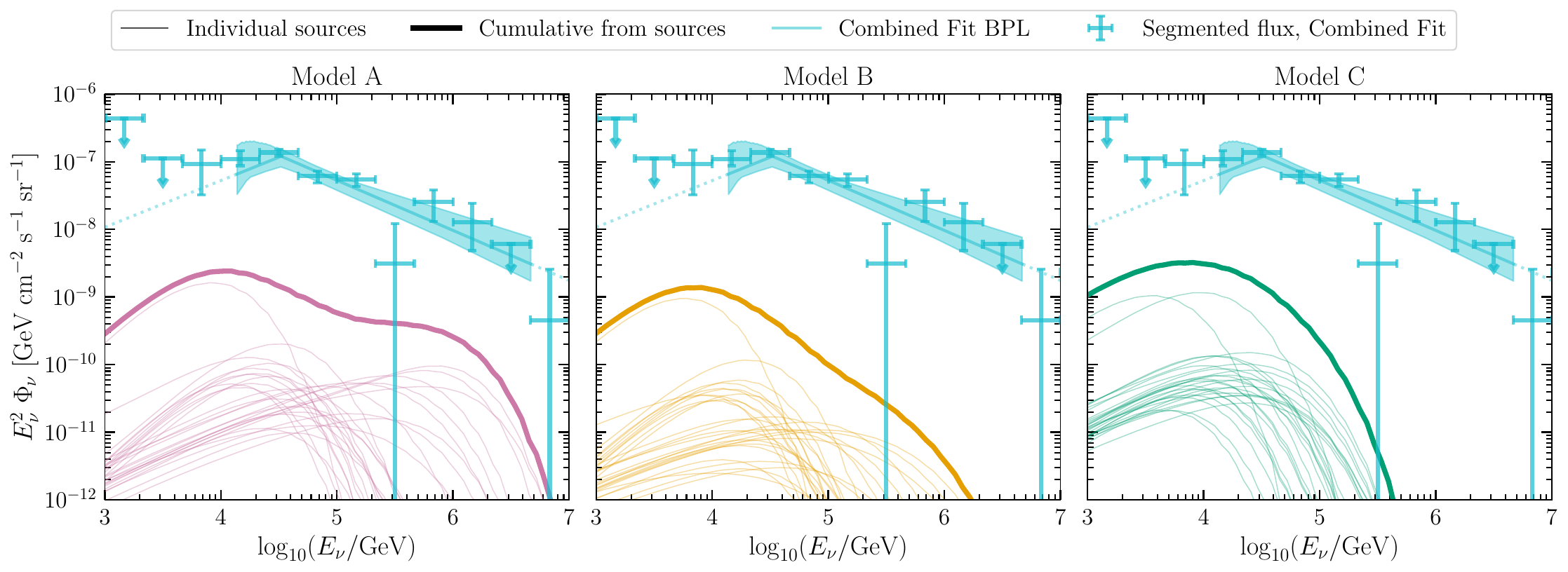}
\caption{\label{fig:summed_diffuse}Predicted all-sky averaged all-flavor neutrino energy fluxes from the X-HiNAC sample for Models A (left), B (middle), and C (right). The thin curves show the contributions from individual AGNs in Table~\ref{tab:swift_bat_sample}, while the thick curves represent their sum. For Models A and B, motivated by the IceCube observations of the all-sky flux, we choose $\mathcal{R} = 30$ for $\lambda_{\rm Edd}<0.3$, while for Model C we have $\mathcal{R} = 10$ irrespective of $\lambda_{\rm Edd}$. For comparison, the measured all-sky astrophysical neutrino flux and its uncertainty are also shown~\citep{IceCube:2025tgp}. The light blue shaded regions show the $68\%$ C.L. uncertainties of the IceCube broken power-law (BPL) fits from the Combined Fit analyses with the corresponding best-fit spectra shown as dotted lines. The segmented flux measurements from the Combined Fit analysis are shown by the light blue crosses.
}
\end{figure*}
In this work, we have constructed X-HiNAC, a sample of 28 intrinsic X-ray bright AGNs as promising targets for high-energy neutrino searches with current and next-generation neutrino telescopes. Starting from the {\em Swift}/BAT and BASS surveys, we select sources with intrinsic 2–195 keV X-ray fluxes exceeding $10^{-7} \mathrm{GeV cm^{-2} s^{-1}}$, motivated by the projected discovery potential of IceCube-Gen2. We further exclude AGNs with prominent relativistic jets and apply additional source specific validation criteria to obtain our final sample.

Combining information from these surveys with dedicated {\em NuSTAR} observations and other measurements available in the literature, we have compiled their intrinsic X-ray spectra
, luminosity distances, and SMBH masses (Table~\ref{tab:swift_bat_sample}). Using the magnetically powered corona model, we have calculated the expected neutrino emission from these sources under three representative model assumptions and investigated their detection prospects with IceCube-Gen2 and KM3NeT. Our primary goal is to provide a systematically characterized sample that can serve as a reference for future individual source and stacking searches.

Our results demonstrate that several nearby X-ray bright AGNs are promising targets for future neutrino observations, although their detectability depends on the details of the model and detector responses. The predicted neutrino spectra and corresponding event distributions illustrate the substantial differences among individual sources and between IceCube-Gen2 and KM3NeT. In particular, NGC 1068 and Circinus Galaxy emerge as promising individual targets for IceCube-Gen2 and KM3NeT, respectively, with projected detection significances exceeding $5\sigma$ under favorable model assumptions. While only a subset of sources may yield significant individual detections, combining their contributions can substantially improve the discovery prospects. In particular, our approximate stacking analysis, excluding NGC 1068, indicates that a collective signal could reach a nominal significance of approximately $5\sigma$ within $\sim 6 - 8$ years for Models A and C, whereas Model B generally requires longer exposures (see Figure~\ref{fig:sig}). These projections motivate dedicated stacking analyses.

To place our results in the broader context of the astrophysical neutrino flux, Figure~\ref{fig:summed_diffuse}
compares the predicted all-sky averaged neutrino flux from our sample with the all-sky flux measured by IceCube~\citep{IceCube:2025tgp}. The combined all-flavor neutrino flux reaches $E_\nu^2 \Phi_\nu \sim (1 - 3) \times 10^{-9}\ {\rm GeV\ cm^{-2} s^{-1} sr^{-1}}$, corresponding to approximately $1 - 3\%$ of the measured all-sky flux intensity at $\sim 10$ TeV. Model A predicts a broader contribution extending to higher energies, whereas Models B and C are concentrated predominantly below $\sim 100\ {\rm TeV}$. Since our sample contains only a selected sample of nearby X-ray bright AGNs, these estimates do not represent the contribution from the entire AGN population. Most of the contribution originates from AGNs at $z\sim1-2$~\citep{Murase:2019vdl}. Nevertheless, the relatively modest contribution to the all-sky flux intensity, together with the promising individual source and stacking prospects, demonstrates that targeted neutrino searches can be sensitive to source populations that contribute only a fraction of the astrophysical neutrino flux. Contributions from the AGN population were discussed in \cite{Carpio:2026xkf}, where they assumed $R=10R_S$ (Model II), showing that an AGN population with parameters similar to NGC 1068, Model A or B can account for the diffuse neutrino flux in the 1 TeV $-$ 20 TeV range.

An important challenge in constructing this sample is the substantial uncertainty in several observationally inferred AGN properties. In particular, estimates of the SMBH mass can differ by up to an order of magnitude between different measurement techniques, while uncertainties in the hydrogen column density and intrinsic X-ray luminosity, especially for heavily obscured sources, can introduce additional systematic uncertainties. Luminosity distances can also be uncertain for nearby AGNs, where peculiar velocities may significantly affect redshift-based estimates. Furthermore, source identification and classification are not always unambiguous across different surveys.
For example, NGC 1275 and Cen A were initially flagged as Seyfert galaxies in the initial 70-month BASS catalog~\citep{Koss:2017ilk}. However, NGC 1275 is no longer classified as a Seyfert in the updated BASS catalog~\citep{Koss:2022lkb}, while the classification of Cen A remains ambiguous due to its prominent jet activity~\citep{2011A&A...531A..70B}, despite its optical classification as a Seyfert. Such ambiguities highlight the importance of carefully distinguishing between coronal and jet-dominated AGNs when constructing samples for neutrino searches.
Together, these observational and classification uncertainties can affect the sample selection and the predicted neutrino emission. Our study therefore highlights the importance of improved SMBH mass measurements, broadband X-ray spectroscopy, and consistent source characterization for future multimessenger investigations.

The precise X-ray flux tiers or rankings in Table~\ref{tab:swift_bat_sample} should be interpreted with caution. As discussed in Section~\ref{sec:cat}, the inferred intrinsic X-ray fluxes depend on the adopted observations and spectral modeling, particularly for heavily obscured AGNs. Moreover, several sources beyond the brightest few have comparable fluxes, such that variations by factors of $\sim 2 – 3$, arising from differences in spectral analyses or intrinsic source variability, can alter their relative rankings. Since AGN X-ray emission can also vary substantially over decade timescales, these rankings may evolve during the IceCube-Gen2 era. To alleviate this problem, we also classify the sources in the X-HiNAC sample into broadly comparable intrinsic X-ray flux based tiers (see Section~\ref{subsec:rank_tiers}). However, even the Tier I sources in the sample can have considerable variations in the intrinsic X-ray flux affecting corresponding neutrino fluxes. Furthermore, as illustrated by the source-dependent neutrino spectra and detection prospects in Section~\ref{sec:res}, X-ray brightness alone does not uniquely determine neutrino detectability, which also depends on the underlying model scenario and detector response. Thus, our sample provides a reference sample of promising neutrino targets rather than a definitive ordering, and future searches will benefit from utilizing updated X-ray measurements and source-specific neutrino modeling.

We emphasize that Models A, B, and C used in this work are representative scenarios motivated by the all-sky neutrino background spectrum, rather than models optimized for individual AGNs. In particular, our calculations do not account for the source-to-source dispersion in physical parameters governing neutrino production, such as the coronal size, turbulence, and CR acceleration efficiency. Consequently, the predicted neutrino spectra and detection significances should be interpreted as illustrative benchmarks rather than definitive source-specific forecasts. Nevertheless, simple power-law spectra or a source-independent spectral template may not adequately capture the diversity of neutrino emission from AGN coronae. Physically motivated spectral energy distribution (SED) templates, such as those provided by the magnetically powered corona model, offer a more realistic framework for future catalog-based searches, particularly when combined with source-dependent parameters and physically motivated priors.

Looking ahead, improved broadband X-ray observations and SMBH mass measurements, together with source-dependent neutrino emission models incorporating physically motivated priors on coronal parameters, will be essential for reducing theoretical and observational uncertainties. The increasing sensitivity and complementary sky coverage of IceCube-Gen2 and upcoming neutrino telescopes in the Northern Hemisphere will provide opportunities to test these predictions through both individual source and stacking searches, with P-ONE~\citep{P-ONE:2020ljt} offering detection prospects comparable to those of KM3NeT. Additionally, other current and upcoming facilities, including Baikal-GVD~\citep{Baikal-GVD:2018isr} and TRIDENT~\citep{TRIDENT:2022hql}, will further expand the capabilities of high-energy neutrino observations. Although our detection forecasts are based on a specific corona model, X-HiNAC can be used to investigate alternative neutrino production scenarios and guide future searches. 
Ultimately, combining improved AGN characterization with high-energy neutrino observations will help establish the role of accreting SMBHs as CR accelerators and contributors to the high-energy neutrino sky.
\begin{acknowledgments}
M.\,M thanks Daniele Rogantini for useful discussions. K.\,M. also thanks Paolo Coppi for the discussion. 
M.\,M and K.\,M. acknowledge funding from NSF Grant No.~AST-2108466, through which this work was proposed and initiated. M.\,M acknowledges support from the FermiForward Discovery Group, LLC under Contract No. 89243024CSC000002 with the U.S. Department of Energy, Office of Science, Office of High Energy Physics. M.\,M. also acknowledges the support of the NSF-Simons AI-Institute for the Sky (SkAI) via grants NSF AST-2421845 and Simons Foundation MPS-AI-00010513. J.C. acknowledges support from the Carlsberg Foundation, grant CF25-1982. H.\,P. and K.\,M. thank support from Penn State for supporting undergraduate research. A.K. acknowledges support from NASA award 80NSSC23M0104. 
K.M. also acknowledges the NSF Grant No.~AST-2308021 and Mamoru Yanagisawa for his generous donation and continuous support throughout this project. 
\end{acknowledgments}
\bibliography{refs}{}

\begin{thebibliography}{}
\expandafter\ifx\csname natexlab\endcsname\relax\def\natexlab#1{#1}\fi
\providecommand{\url}[1]{\href{#1}{#1}}
\providecommand{\dodoi}[1]{doi:~\href{http://doi.org/#1}{\nolinkurl{#1}}}
\providecommand{\doeprint}[1]{\href{http://ascl.net/#1}{\nolinkurl{http://ascl.net/#1}}}
\providecommand{\doarXiv}[1]{\href{https://arxiv.org/abs/#1}{\nolinkurl{https://arxiv.org/abs/#1}}}

\bibitem[{CMS(2011)}]{CMS-NOTE-2011-005}
 2011, {Procedure for the LHC Higgs boson search combination in Summer 2011}, Tech. Rep. CMS-NOTE-2011-005. ATL-PHYS-PUB-2011-11, CERN, Geneva.
\newblock \url{http://cds.cern.ch/record/1379837}

\bibitem[{M.~G. Aartsen {et~al.}(2013{\natexlab{a}})Aartsen {et~al.}}]{IceCube:2013cdw}
Aartsen, M.~G., {et~al.} 2013{\natexlab{a}}, \bibinfo{title}{{First observation of PeV-energy neutrinos with IceCube},} Phys. Rev. Lett., 111, 021103, \dodoi{10.1103/PhysRevLett.111.021103}

\bibitem[{M.~G. Aartsen {et~al.}(2013{\natexlab{b}})Aartsen {et~al.}}]{IceCube:2013low}
Aartsen, M.~G., {et~al.} 2013{\natexlab{b}}, \bibinfo{title}{{Evidence for High-Energy Extraterrestrial Neutrinos at the IceCube Detector},} Science, 342, 1242856, \dodoi{10.1126/science.1242856}

\bibitem[{M.~G. Aartsen {et~al.}(2021)Aartsen {et~al.}}]{IceCube-Gen2:2020qha}
Aartsen, M.~G., {et~al.} 2021, \bibinfo{title}{{IceCube-Gen2: the window to the extreme Universe},} J. Phys. G, 48, 060501, \dodoi{10.1088/1361-6471/abbd48}

\bibitem[{R. Abbasi {et~al.}(2021)Abbasi {et~al.}}]{IceCube:2021xar}
Abbasi, R., {et~al.} 2021, \bibinfo{title}{{IceCube Data for Neutrino Point-Source Searches Years 2008-2018},} \dodoi{10.21234/CPKQ-K003}

\bibitem[{R. Abbasi {et~al.}(2022)Abbasi {et~al.}}]{IceCube:2022der}
Abbasi, R., {et~al.} 2022, \bibinfo{title}{{Evidence for neutrino emission from the nearby active galaxy NGC 1068},} Science, 378, 538, \dodoi{10.1126/science.abg3395}

\bibitem[{R. Abbasi {et~al.}(2023)Abbasi {et~al.}}]{IceCube:2023ame}
Abbasi, R., {et~al.} 2023, \bibinfo{title}{{Observation of high-energy neutrinos from the Galactic plane},} Science, 380, adc9818, \dodoi{10.1126/science.adc9818}

\bibitem[{R. Abbasi {et~al.}(2025{\natexlab{a}})Abbasi {et~al.}}]{IceCube:2024ayt}
Abbasi, R., {et~al.} 2025{\natexlab{a}}, \bibinfo{title}{{Search for Neutrino Emission from Hard X-Ray AGN with IceCube},} Astrophys. J., 981, 131, \dodoi{10.3847/1538-4357/ada94b}

\bibitem[{R. Abbasi {et~al.}(2025{\natexlab{b}})Abbasi {et~al.}}]{IceCube:2025lev}
Abbasi, R., {et~al.} 2025{\natexlab{b}}, \bibinfo{title}{{All-sky Neutrino Point-source Search with IceCube Combined Track and Cascade Data},} Astrophys. J., 995, 11, \dodoi{10.3847/1538-4357/ae113f}

\bibitem[{R. Abbasi {et~al.}(2026{\natexlab{a}})Abbasi {et~al.}}]{IceCube:2026plr}
Abbasi, R., {et~al.} 2026{\natexlab{a}}, \bibinfo{title}{{High-energy neutrino emission from the Milky Way},} \doarXiv{2607.25966}

\bibitem[{R. Abbasi {et~al.}(2026{\natexlab{b}})Abbasi {et~al.}}]{IceCube:2025gdd}
Abbasi, R., {et~al.} 2026{\natexlab{b}}, \bibinfo{title}{{Evidence for Neutrino Emission from X-Ray-bright Active Galactic Nuclei with IceCube},} Astrophys. J. Lett., 1000, L26, \dodoi{10.3847/2041-8213/ae4aad}

\bibitem[{R. Abbasi {et~al.}(2026{\natexlab{c}})Abbasi {et~al.}}]{Abbasi:2026oux}
Abbasi, R., {et~al.} 2026{\natexlab{c}}, \bibinfo{title}{{Evidence for Neutrino Emission from X-Ray Bright Seyfert Galaxies in the Southern Hemisphere Using Enhanced Starting Track Events with IceCube},} Astrophys. J. Lett., 1000, L37, \dodoi{10.3847/2041-8213/ae4aac}

\bibitem[{R. Abbasi {et~al.}(2026{\natexlab{d}})Abbasi {et~al.}}]{IceCube:2025tgp}
Abbasi, R., {et~al.} 2026{\natexlab{d}}, \bibinfo{title}{{Evidence for a Spectral Break or Curvature in the Spectrum of Astrophysical Neutrinos from 5 TeV--10 PeV},} Phys. Rev. Lett., 136, 121002, \dodoi{10.1103/2gh9-d4q7}

\bibitem[{R. Abbasi {et~al.}(2026{\natexlab{e}})Abbasi {et~al.}}]{IceCube:2025zyb}
Abbasi, R., {et~al.} 2026{\natexlab{e}}, \bibinfo{title}{{Time-integrated Southern-sky Neutrino Source Searches with 10 yr of IceCube Starting-track Events at Energies Down to 1 TeV},} Astrophys. J., 998, 37, \dodoi{10.3847/1538-4357/ae2c86}

\bibitem[{R. Abbasi {et~al.}(2026{\natexlab{f}})Abbasi {et~al.}}]{Abbasi:2026ehs}
Abbasi, R., {et~al.} 2026{\natexlab{f}}, \bibinfo{title}{{IceCube Second Track Data Release IceTracks-DR2: Data from 2008-2022 for Neutrino Source Searches},} \doarXiv{2605.19040}

\bibitem[{V.~A. Acciari {et~al.}(2019)Acciari {et~al.}}]{MAGIC:2019fvw}
Acciari, V.~A., {et~al.} 2019, \bibinfo{title}{{Constraints on gamma-ray and neutrino emission from NGC 1068 with the MAGIC telescopes},} Astrophys. J., 883, 135, \dodoi{10.3847/1538-4357/ab3a51}

\bibitem[{S. Adrian-Martinez {et~al.}(2016)Adrian-Martinez {et~al.}}]{KM3Net:2016zxf}
Adrian-Martinez, S., {et~al.} 2016, \bibinfo{title}{{Letter of intent for KM3NeT 2.0},} J. Phys. G, 43, 084001, \dodoi{10.1088/0954-3899/43/8/084001}

\bibitem[{M. Agostini {et~al.}(2020)Agostini {et~al.}}]{P-ONE:2020ljt}
Agostini, M., {et~al.} 2020, \bibinfo{title}{{The Pacific Ocean Neutrino Experiment},} Nature Astron., 4, 913, \dodoi{10.1038/s41550-020-1182-4}

\bibitem[{S. Aiello {et~al.}(2024)Aiello {et~al.}}]{KM3NeT:2024paj}
Aiello, S., {et~al.} 2024, \bibinfo{title}{{Astronomy potential of KM3NeT/ARCA},} Eur. Phys. J. C, 84, 885, \dodoi{10.1140/epjc/s10052-024-13137-2}

\bibitem[{M. Ajello {et~al.}(2023)Ajello, Murase, \& McDaniel}]{Ajello:2023hkh}
Ajello, M., Murase, K., \& McDaniel, A. 2023, \bibinfo{title}{{Disentangling the Hadronic Components in NGC 1068},} Astrophys. J. Lett., 954, L49, \dodoi{10.3847/2041-8213/acf296}

\bibitem[{J. Aleksić {et~al.}(2014)Aleksić, Ansoldi, Antonelli, Antoranz, Babic, Bangale, Barres~de Almeida, Barrio, Becerra~González, Bednarek, Berger, Bernardini, Biland, Blanch, Bock, Bonnefoy, Bonnoli, Borracci, Bretz, Carmona, Carosi, Carreto~Fidalgo, Colin, Colombo, Contreras, Cortina, Covino, Da~Vela, Dazzi, De~Angelis, De~Caneva, De~Lotto, Delgado~Mendez, Doert, Domínguez, Dominis~Prester, Dorner, Doro, Einecke, Eisenacher, Elsaesser, Farina, Ferenc, Fonseca, Font, Frantzen, Fruck, García~López, Garczarczyk, Garrido~Terrats, Gaug, Giavitto, Godinović, González~Muñoz, Gozzini, Hadamek, Hadasch, Herrero, Hildebrand, Hose, Hrupec, Idec, Kadenius, Kellermann, Knoetig, Krause, Kushida, La~Barbera, Lelas, Lewandowska, Lindfors, Lombardi, López, López-Coto, López-Oramas, Lorenz, Lozano, Makariev, Mallot, Maneva, Mankuzhiyil, Mannheim, Maraschi, Marcote, Mariotti, Martínez, Mazin, Menzel, Meucci, Miranda, Mirzoyan, Moralejo, Munar-Adrover, Nakajima, Niedzwiecki, Nilsson, Nowak, Orito, Overkemping, Paiano, Palatiello, Paneque, Paoletti, Paredes, Paredes-Fortuny, Partini, Persic, Prada, Prada~Moroni, Prandini, Preziuso, Puljak, Reinthal, Rhode, Ribó, Rico, Rodriguez~Garcia, Rügamer, Saggion, Saito, Saito, Salvati, Satalecka, Scalzotto, Scapin, Schultz, Schweizer, Shore, Sillanpää, Sitarek, Snidaric, Sobczynska, Spanier, Stamatescu, Stamerra, Steinbring, Storz, Sun, Surić, Takalo, Tavecchio, Terzić, Tescaro, Teshima, Thaele, Tibolla, Torres, Toyama, Treves, Uellenbeck, Vogler, Wagner, Zandanel, Zanin, Balmaverde, Kataoka, Rekola, \& Takahashi}]{Aleksi__2014}
Aleksić, J., Ansoldi, S., Antonelli, L.~A., {et~al.} 2014, \bibinfo{title}{Contemporaneous observations of the radio galaxy NGC 1275 from radio to very high energy<i>γ</i>-rays,} Astronomy \& Astrophysics, 564, A5, \dodoi{10.1051/0004-6361/201322951}

\bibitem[{D.~E. Alexandreas {et~al.}(1993)Alexandreas {et~al.}}]{Alexandreas:1992ek}
Alexandreas, D.~E., {et~al.} 1993, \bibinfo{title}{{Point source search techniques in ultrahigh-energy gamma-ray astronomy},} Nucl. Instrum. Meth. A, 328, 570, \dodoi{10.1016/0168-9002(93)90677-A}

\bibitem[{A. Ambrosone(2024)Ambrosone}]{Ambrosone:2024zrf}
Ambrosone, A. 2024, \bibinfo{title}{{Berezinsky hidden sources: an emergent tension in the high-energy neutrino sky?},} JCAP, 09, 075, \dodoi{10.1088/1475-7516/2024/09/075}

\bibitem[{A. Amorim {et~al.}(2021)Amorim {et~al.}}]{GRAVITY:2021ory}
Amorim, A., {et~al.} 2021, \bibinfo{title}{{A geometric distance to the supermassive black Hole of NGC 3783},} Astron. Astrophys., 654, A85, \dodoi{10.1051/0004-6361/202141426}

\bibitem[{P. Ar{\'e}valo {et~al.}(2014)Ar{\'e}valo {et~al.}}]{Arevalo:2014zva}
Ar{\'e}valo, P., {et~al.} 2014, \bibinfo{title}{{The 2-79 keV X-ray Spectrum of the Circinus Galaxy with NuSTAR, XMM-Newton and Chandra: a Fully Compton-Thick Active Galactic Nucleus},} Astrophys. J., 791, 81, \dodoi{10.1088/0004-637X/791/2/81}

\bibitem[{A.~D. Avrorin {et~al.}(2018)Avrorin {et~al.}}]{Baikal-GVD:2018isr}
Avrorin, A.~D., {et~al.} 2018, \bibinfo{title}{{Baikal-GVD: status and prospects},} EPJ Web Conf., 191, 01006, \dodoi{10.1051/epjconf/201819101006}

\bibitem[{M. Balokovi{\'c} {et~al.}(2018)Balokovi{\'c}, Brightman, Harrison, Comastri, Ricci, Buchner, Gandhi, Farrah, \& Stern}]{Balokovic:2018bcm}
Balokovi{\'c}, M., Brightman, M., Harrison, F.~A., {et~al.} 2018, \bibinfo{title}{{New Spectral Model for Constraining Torus Covering Factors from Broadband X-ray Spectra of Active Galactic Nuclei},} Astrophys. J., 854, 42, \dodoi{10.3847/1538-4357/aaa7eb}

\bibitem[{W.~H. {Baumgartner} {et~al.}(2013){Baumgartner}, {Tueller}, {Markwardt}, {Skinner}, {Barthelmy}, {Mushotzky}, {Evans}, \& {Gehrels}}]{2013ApJS..207...19B}
{Baumgartner}, W.~H., {Tueller}, J., {Markwardt}, C.~B., {et~al.} 2013, \bibinfo{title}{{The 70 Month Swift-BAT All-sky Hard X-Ray Survey},} \apjs, 207, 19, \dodoi{10.1088/0067-0049/207/2/19}

\bibitem[{V. {Beckmann} {et~al.}(2011){Beckmann}, {Jean}, {Lubi{\'n}ski}, {Soldi}, \& {Terrier}}]{2011A&A...531A..70B}
{Beckmann}, V., {Jean}, P., {Lubi{\'n}ski}, P., {Soldi}, S., \& {Terrier}, R. 2011, \bibinfo{title}{{The hard X-ray emission of Centaurus A},} \aap, 531, A70, \dodoi{10.1051/0004-6361/201016020}

\bibitem[{M.~C. Bentz {et~al.}(2023)Bentz, Markham, Rosborough, Onken, Street, Valluri, \& Treu}]{bentz2023velocityresolvedreverberationmappingngc}
Bentz, M.~C., Markham, M., Rosborough, S., {et~al.} 2023, Velocity-Resolved Reverberation Mapping of NGC 3227, \doarXiv{2310.03904}

\bibitem[{M.~C. {Bentz} {et~al.}(2022){Bentz}, {Williams}, \& {Treu}}]{2022ApJ...934..168B}
{Bentz}, M.~C., {Williams}, P.~R., \& {Treu}, T. 2022, \bibinfo{title}{{The Broad Line Region and Black Hole Mass of NGC 4151},} \apj, 934, 168, \dodoi{10.3847/1538-4357/ac7c0a}

\bibitem[{C. Blanco {et~al.}(2025)Blanco, Hooper, Linden, \& Pinetti}]{Blanco:2023dfp}
Blanco, C., Hooper, D., Linden, T., \& Pinetti, E. 2025, \bibinfo{title}{{Neutrino and gamma-ray emissions from NGC 1068},} Phys. Rev. D, 112, 123016, \dodoi{10.1103/wnjh-7nwp}

\bibitem[{M. {Brightman} \& K. {Nandra}(2011){Brightman} \& {Nandra}}]{Brightman:2011}
{Brightman}, M., \& {Nandra}, K. 2011, \bibinfo{title}{{An XMM-Newton spectral survey of 12 {\ensuremath{\mu}}m selected galaxies - I. X-ray data},} \mnras, 413, 1206, \dodoi{10.1111/j.1365-2966.2011.18207.x}

\bibitem[{J.~A. Carpio {et~al.}(2026)Carpio, Kheirandish, \& Murase}]{Carpio:2026xkf}
Carpio, J.~A., Kheirandish, A., \& Murase, K. 2026, \bibinfo{title}{{Multimessenger Characterization of High-energy Neutrino Emission from the Brightest Neutrino-active Galactic Nuclei},} Astrophys. J., 1005, 61, \dodoi{10.3847/1538-4357/ae7071}

\bibitem[{S. Chalise {et~al.}(2022)Chalise, Lohfink, Chauhan, Russell, Buisson, \& Mallick}]{10.1093/mnras/stac2953}
Chalise, S., Lohfink, A.~M., Chauhan, J., {et~al.} 2022, \bibinfo{title}{Broad-band spectral analysis of Mrk 926 using multi-epoch X-ray observations,} Monthly Notices of the Royal Astronomical Society, 517, 4788, \dodoi{10.1093/mnras/stac2953}

\bibitem[{R. Chatterjee {et~al.}(2009)Chatterjee, Marscher, Jorstad, Olmstead, McHardy, Aller, Aller, Lähteenmäki, Tornikoski, Hovatta, Marshall, Miller, Ryle, Chicka, Benker, Bottorff, Brokofsky, Campbell, Chonis, Gaskell, Gaynullina, Grankin, Hedrick, Ibrahimov, Klimek, Kruse, Masatoshi, Miller, Pan, Petersen, Peterson, Shen, Strel’nikov, Tao, Watkins, \& Wheeler}]{Chatterjee_2009}
Chatterjee, R., Marscher, A.~P., Jorstad, S.~G., {et~al.} 2009, \bibinfo{title}{DISK–JET CONNECTION IN THE RADIO GALAXY 3C 120,} The Astrophysical Journal, 704, 1689–1703, \dodoi{10.1088/0004-637x/704/2/1689}

\bibitem[{M. Chernyakova {et~al.}(2004)Chernyakova, Favre, Courvoisier, Lutovinov, Molkov, Beckmann, Gros, Gehrels, Produit, Walter, \& Zdziarski}]{chernyakova2004integralrxteobservationsbroadline}
Chernyakova, M., Favre, P., Courvoisier, T. J.~L., {et~al.} 2004, INTEGRAL and RXTE observations of broad-line radio galaxy 3C 111, \doarXiv{astro-ph/0404046}

\bibitem[{A. Das {et~al.}(2026{\natexlab{a}})Das, Feng, Young, Pandey, Kimura, \& Murase}]{Das:2026aed}
Das, A., Feng, Q., Young, E., {et~al.} 2026{\natexlab{a}}, \bibinfo{title}{{Interpreting Swift and NuSTAR Observations of the Low-luminosity Active Galactic Nucleus NGC 4278 with Radiatively Inefficient Accretion Flows and Implications for Neutrino Emission},} Astrophys. J., 1006, 186, \dodoi{10.3847/1538-4357/ae7e7f}

\bibitem[{A. Das {et~al.}(2026{\natexlab{b}})Das, Murase, \& Zhang}]{Das:2026rwt}
Das, A., Murase, K., \& Zhang, B.~T. 2026{\natexlab{b}}, \bibinfo{title}{{Multimessenger Constraints on Production Sites of High-energy Neutrinos from NGC 1068},} Astrophys. J. Lett., 1009, L27, \dodoi{10.3847/2041-8213/ae9ba3}

\bibitem[{A. Das {et~al.}(2024)Das, Zhang, \& Murase}]{Das_2024}
Das, A., Zhang, B.~T., \& Murase, K. 2024, \bibinfo{title}{Revealing the Production Mechanism of High-energy Neutrinos from NGC 1068,} The Astrophysical Journal, 972, 44, \dodoi{10.3847/1538-4357/ad5a04}

\bibitem[{N. Das {et~al.}(2025)Das, Bentz, Vasiliev, Valluri, Onken, Raimundo, \& Vestergaard}]{das2025stellardynamicalmasscentral}
Das, N., Bentz, M.~C., Vasiliev, E., {et~al.} 2025, A Stellar Dynamical Mass for the Central Black Hole in MCG$-$06-30-15, \doarXiv{2509.01017}

\bibitem[{M. {Durr{\'e}} \& J. {Mould}(2019){Durr{\'e}} \& {Mould}}]{2019ApJ...870...37D}
{Durr{\'e}}, M., \& {Mould}, J. 2019, \bibinfo{title}{{The AGN Ionization Cones of NGC 5728. II. Kinematics},} \apj, 870, 37, \dodoi{10.3847/1538-4357/aaf000}

\bibitem[{H. {Engel} {et~al.}(2010){Engel}, {Davies}, {Genzel}, {Tacconi}, {Hicks}, {Sturm}, {Naab}, {Johansson}, {Karl}, {Max}, {Medling}, \& {van der Werf}}]{2010A&A...524A..56E}
{Engel}, H., {Davies}, R.~I., {Genzel}, R., {et~al.} 2010, \bibinfo{title}{{NGC 6240: merger-induced star formation and gas dynamics},} \aap, 524, A56, \dodoi{10.1051/0004-6361/201015338}

\bibitem[{H.-C. {Feng} {et~al.}(2024){Feng}, {Li}, {Bai}, {Liu}, {Lu}, {Pang}, {Sun}, {Wang}, {Zhang}, \& {Zhou}}]{2024ApJ...976..176F}
{Feng}, H.-C., {Li}, S.-S., {Bai}, J.~M., {et~al.} 2024, \bibinfo{title}{{Velocity-resolved Reverberation Mapping of Changing-look Active Galactic Nucleus NGC 4151 during Outburst Stage. II. Results of Four Seasons of Observation},} \apj, 976, 176, \dodoi{10.3847/1538-4357/ad8568}

\bibitem[{H.-C. {Feng} {et~al.}(2025){Feng}, {Li}, {Bai}, {Liu}, {Lu}, {Pang}, {Sun}, {Wang}, {Xu}, {Zhang}, \& {Zhou}}]{2025ApJ...979..131F}
{Feng}, H.-C., {Li}, S.-S., {Bai}, J.~M., {et~al.} 2025, \bibinfo{title}{{Reverberation Mapping of Two Variable Active Galactic Nuclei: Probing the Distinct Characteristics of the Inner and Outer Broad-line Regions},} \apj, 979, 131, \dodoi{10.3847/1538-4357/ad9c71}

\bibitem[{D.~F.~G. Fiorillo {et~al.}(2024)Fiorillo, Petropoulou, Comisso, Peretti, \& Sironi}]{Fiorillo:2023dts}
Fiorillo, D. F.~G., Petropoulou, M., Comisso, L., Peretti, E., \& Sironi, L. 2024, \bibinfo{title}{{TeV Neutrinos and Hard X-Rays from Relativistic Reconnection in the Corona of NGC 1068},} Astrophys. J., 961, L14, \dodoi{10.3847/2041-8213/ad192b}

\bibitem[{J.~F. {Gallimore} {et~al.}(2024){Gallimore}, {Impellizzeri}, {Aghelpasand}, {Gao}, {Hostetter}, \& {Lankhaar}}]{2024ApJ...975L...9G}
{Gallimore}, J.~F., {Impellizzeri}, C.~M.~V., {Aghelpasand}, S., {et~al.} 2024, \bibinfo{title}{{The Discovery of Polarized Water Vapor Megamaser Emission in a Molecular Accretion Disk},} \apjl, 975, L9, \dodoi{10.3847/2041-8213/ad864f}

\bibitem[{J. Gerssen {et~al.}(2004)Gerssen, van~der Marel, Axon, Mihos, Hernquist, \& Barnes}]{Gerssen_2004}
Gerssen, J., van~der Marel, R.~P., Axon, D., {et~al.} 2004, \bibinfo{title}{<i>Hubble Space Telescope</i>Observations of NGC 6240: A Case Study of an Ultraluminous Infrared Galaxy with Obscured Activity,} The Astronomical Journal, 127, 75–89, \dodoi{10.1086/380223}

\bibitem[{V.~E. Gianolli {et~al.}(2023)Gianolli {et~al.}}]{Gianolli:2023zji}
Gianolli, V.~E., {et~al.} 2023, \bibinfo{title}{{Uncovering the geometry of the hot X-ray corona in the Seyfert galaxy NGC 4151 with IXPE},} Mon. Not. Roy. Astron. Soc., 523, 4468, \dodoi{10.1093/mnras/stad1697}

\bibitem[{M. Gliozzi {et~al.}(2024)Gliozzi, Williams, Akylas, Papadakis, Shuvo, Halavatkar, \& Alt}]{10.1093/mnras/stad3974}
Gliozzi, M., Williams, J.~K., Akylas, A., {et~al.} 2024, \bibinfo{title}{Comparing indirect methods for black hole masses in AGN: the good, the bad, and the ugly,} Monthly Notices of the Royal Astronomical Society, 528, 3417, \dodoi{10.1093/mnras/stad3974}

\bibitem[{A.~G. {Gonzalez} {et~al.}(2025){Gonzalez}, {Gallo}, {Ingram}, {Xu}, \& {Buhariwalla}}]{2025ApJ...994..151G}
{Gonzalez}, A.~G., {Gallo}, L.~C., {Ingram}, A., {Xu}, Y., \& {Buhariwalla}, M.~Z. 2025, \bibinfo{title}{{Spectrotemporal Analysis of X-Ray Reverberation in the Seyfert 1.5 Galaxy 4U 1344─60},} \apj, 994, 151, \dodoi{10.3847/1538-4357/ae0ca3}

\bibitem[{L.~J. Greenhill {et~al.}(1997)Greenhill, Ellingsen, Norris, Gough, Sinclair, Moran, \& Mushotzky}]{Greenhill:1996gz}
Greenhill, L.~J., Ellingsen, S.~P., Norris, R.~P., {et~al.} 1997, \bibinfo{title}{{Extremely rapid variations of water maser emission from the circinus galaxy},} Astrophys. J. Lett., 474, L103, \dodoi{10.1086/310434}

\bibitem[{L.~J. {Greenhill} {et~al.}(2003){Greenhill}, {Booth}, {Ellingsen}, {Herrnstein}, {Jauncey}, {McCulloch}, {Moran}, {Norris}, {Reynolds}, \& {Tzioumis}}]{2003ApJ...590..162G}
{Greenhill}, L.~J., {Booth}, R.~S., {Ellingsen}, S.~P., {et~al.} 2003, \bibinfo{title}{{A Warped Accretion Disk and Wide-Angle Outflow in the Inner Parsec of the Circinus Galaxy},} \apj, 590, 162, \dodoi{10.1086/374862}

\bibitem[{M. Honda {et~al.}(2015)Honda, Sajjad~Athar, Kajita, Kasahara, \& Midorikawa}]{Honda:2015fha}
Honda, M., Sajjad~Athar, M., Kajita, T., Kasahara, K., \& Midorikawa, S. 2015, \bibinfo{title}{{Atmospheric neutrino flux calculation using the NRLMSISE-00 atmospheric model},} Phys. Rev. D, 92, 023004, \dodoi{10.1103/PhysRevD.92.023004}

\bibitem[{ {IceCube-Gen2 Collaboration}(2024){IceCube-Gen2 Collaboration}}]{Gen2_TDR}
{IceCube-Gen2 Collaboration}. 2024, {IceCube-Gen2 Technical Design Report}, Tech. rep.
\newblock \url{https://icecube-gen2.wisc.edu/science/publications/tdr/}

\bibitem[{S. Inoue {et~al.}(2023)Inoue, Cerruti, Murase, \& Liu}]{Inoue:2022yak}
Inoue, S., Cerruti, M., Murase, K., \& Liu, R.-Y. 2023, \bibinfo{title}{{Multimessenger emission from winds and tori in active galactic nuclei},} PoS, ICRC2023, 1161, \dodoi{10.22323/1.444.1161}

\bibitem[{Y. Inoue {et~al.}(2020)Inoue, Khangulyan, \& Doi}]{Inoue:2019yfs}
Inoue, Y., Khangulyan, D., \& Doi, A. 2020, \bibinfo{title}{{On the Origin of High-energy Neutrinos from NGC 1068: The Role of Nonthermal Coronal Activity},} Astrophys. J. Lett., 891, L33, \dodoi{10.3847/2041-8213/ab7661}

\bibitem[{T. Kawamuro {et~al.}(2016)Kawamuro, Ueda, Tazaki, Ricci, \& Terashima}]{Kawamuro_2016}
Kawamuro, T., Ueda, Y., Tazaki, F., Ricci, C., \& Terashima, Y. 2016, \bibinfo{title}{SUZAKU OBSERVATIONS OF MODERATELY OBSCURED (COMPTON-THIN) ACTIVE GALACTIC NUCLEI SELECTED BY SWIFT/BAT HARD X-RAY SURVEY,} The Astrophysical Journal Supplement Series, 225, 14, \dodoi{10.3847/0067-0049/225/1/14}

\bibitem[{A. Kheirandish {et~al.}(2021)Kheirandish, Murase, \& Kimura}]{Kheirandish:2021wkm}
Kheirandish, A., Murase, K., \& Kimura, S.~S. 2021, \bibinfo{title}{{High-energy Neutrinos from Magnetized Coronae of Active Galactic Nuclei and Prospects for Identification of Seyfert Galaxies and Quasars in Neutrino Telescopes},} Astrophys. J., 922, 45, \dodoi{10.3847/1538-4357/ac1c77}

\bibitem[{G.~A. {Khorunzhev} {et~al.}(2012){Khorunzhev}, {Sazonov}, {Burenin}, \& {Tkachenko}}]{2012AstL...38..475K}
{Khorunzhev}, G.~A., {Sazonov}, S.~Y., {Burenin}, R.~A., \& {Tkachenko}, A.~Y. 2012, \bibinfo{title}{{Masses and accretion rates of supermassive black holes in active galactic nuclei from the INTEGRAL survey},} Astronomy Letters, 38, 475, \dodoi{10.1134/S1063773712080026}

\bibitem[{J. Kormendy \& L.~C. Ho(2013)Kormendy \& Ho}]{Kormendy:2013dxa}
Kormendy, J., \& Ho, L.~C. 2013, \bibinfo{title}{{Coevolution (Or Not) of Supermassive Black Holes and Host Galaxies},} Ann. Rev. Astron. Astrophys., 51, 511, \dodoi{10.1146/annurev-astro-082708-101811}

\bibitem[{M. Koss {et~al.}(2017)Koss {et~al.}}]{Koss:2017ilk}
Koss, M., {et~al.} 2017, \bibinfo{title}{{BAT AGN Spectroscopic Survey I: Spectral Measurements, Derived Quantities, and AGN Demographics},} Astrophys. J., 850, 74, \dodoi{10.3847/1538-4357/aa8ec9}

\bibitem[{M.~J. Koss {et~al.}(2022{\natexlab{a}})Koss {et~al.}}]{Koss:2022lkb}
Koss, M.~J., {et~al.} 2022{\natexlab{a}}, \bibinfo{title}{{BAT AGN Spectroscopic Survey XXI: The Data Release 2 Overview},} Astrophys. J. Suppl., 261, 1, \dodoi{10.3847/1538-4365/ac6c8f}

\bibitem[{M.~J. Koss {et~al.}(2022{\natexlab{b}})Koss {et~al.}}]{Koss:2022qwh}
Koss, M.~J., {et~al.} 2022{\natexlab{b}}, \bibinfo{title}{{BASS XXII: The BASS DR2 AGN Catalog and Data},} Astrophys. J. Suppl., 261, 2, \dodoi{10.3847/1538-4365/ac6c05}

\bibitem[{E. Kun {et~al.}(2024)Kun, Bartos, Becker~Tjus, Biermann, Franckowiak, Halzen, del Palacio, \& Woo}]{Kun:2024meq}
Kun, E., Bartos, I., Becker~Tjus, J., {et~al.} 2024, \bibinfo{title}{{Possible correlation between unabsorbed hard x rays and neutrinos in radio-loud and radio-quiet active galactic nuclei},} Phys. Rev. D, 110, 123014, \dodoi{10.1103/PhysRevD.110.123014}

\bibitem[{C.~Y. {Kuo} {et~al.}(2019){Kuo}, {Braatz}, {Impellizzeri}, {Gao}, {Pesce}, {Reid}, {Condon}, {Kamali}, {Henkel}, \& {Greene}}]{2019arXiv191014314K}
{Kuo}, C.~Y., {Braatz}, J.~A., {Impellizzeri}, C.~M.~V., {et~al.} 2019, \bibinfo{title}{{The Megamaser Cosmology Project. XII. VLBI Imaging of H$_{2}$O Maser Emission in Three Active Galaxies and the Effect of AGN Winds on Disk Dynamics},} arXiv e-prints, arXiv:1910.14314, \dodoi{10.48550/arXiv.1910.14314}

\bibitem[{S. Laha {et~al.}(2014)Laha, Guainazzi, Dewangan, Chakravorty, \& Kembhavi}]{Laha:2014faa}
Laha, S., Guainazzi, M., Dewangan, G.~C., Chakravorty, S., \& Kembhavi, A.~K. 2014, \bibinfo{title}{{Warm absorbers in X-rays (WAX), a comprehensive high-resolution grating spectral study of a sample of Seyfert galaxies {\textendash} I. A global view and frequency of occurrence of warm absorbers.},} Mon. Not. Roy. Astron. Soc., 441, 2613, \dodoi{10.1093/mnras/stu669}

\bibitem[{S. Le~Bihan {et~al.}(2026)Le~Bihan, Lemoine, \& Rieger}]{LeBihan:2026rqq}
Le~Bihan, S., Lemoine, M., \& Rieger, F. 2026, \bibinfo{title}{{Self-consistent modelling of neutrino production in turbulent black hole coronae},} Astron. Astrophys., 712, A218, \dodoi{10.1051/0004-6361/202660090}

\bibitem[{K.~M. Leighly {et~al.}(1995)Leighly, Dietrich, Waltman, Edelson, George, Malkan, Matsuoka, Mushotzky, \& Peterson}]{leighly1995resultsmonitoringbroadlineradio}
Leighly, K.~M., Dietrich, M., Waltman, E., {et~al.} 1995, Results from Monitoring the Broad-Line Radio Galaxy 3C 390.3, \doarXiv{astro-ph/9512059}

\bibitem[{M. Lemoine \& F. Rieger(2025)Lemoine \& Rieger}]{Lemoine:2024roa}
Lemoine, M., \& Rieger, F. 2025, \bibinfo{title}{{Neutrinos from stochastic acceleration in black hole environments},} Astron. Astrophys., 697, A124, \dodoi{10.1051/0004-6361/202453296}

\bibitem[{Y.-R. {Li} {et~al.}(2024){Li}, {Hu}, {Yao}, {Chen}, {Bai}, {Yang}, {Du}, {Fang}, {Fu}, {Liu}, {Peng}, {Songsheng}, {Wang}, {Xiao}, {Zhai}, {Winkler}, {Bai}, {Ho}, {Petrov}, {Aceituno}, \& {Wang}}]{2024ApJ...974...86L}
{Li}, Y.-R., {Hu}, C., {Yao}, Z.-H., {et~al.} 2024, \bibinfo{title}{{Spectroastrometry and Reverberation Mapping of Active Galactic Nuclei. I. The H{\ensuremath{\beta}} Broad-line Region Structure and Black Hole Masses of Five Quasars},} \apj, 974, 86, \dodoi{10.3847/1538-4357/ad6906}

\bibitem[{A.~Y. Lien {et~al.}(2025)Lien {et~al.}}]{Lien:2025zzo}
Lien, A.~Y., {et~al.} 2025, \bibinfo{title}{{The 157 Month Swift/BAT All-sky Hard X-Ray Survey},} Astrophys. J., 989, 161, \dodoi{10.3847/1538-4357/ade676}

\bibitem[{P. Lira {et~al.}(2011)Lira, Arévalo, Uttley, McHardy, \& Breedt}]{Lira_2011}
Lira, P., Arévalo, P., Uttley, P., McHardy, I., \& Breedt, E. 2011, \bibinfo{title}{Optical and near-IR long-term monitoring of NGC 3783 and MR 2251−178: evidence for variable near-IR emission from thin accretion discs: Optical and near-IR monitoring,} Monthly Notices of the Royal Astronomical Society, 415, 1290–1303, \dodoi{10.1111/j.1365-2966.2011.18774.x}

\bibitem[{G. Lodato \& G. Bertin(2003)Lodato \& Bertin}]{Lodato:2002cv}
Lodato, G., \& Bertin, G. 2003, \bibinfo{title}{{Non-Keplerian rotation in the nucleus of NGC 1068: Evidence for a massive accretion disk?},} Astron. Astrophys., 398, 517, \dodoi{10.1051/0004-6361:20021672}

\bibitem[{S. Marchesi {et~al.}(2018)Marchesi, Ajello, Marcotulli, Comastri, Lanzuisi, \& Vignali}]{Marchesi:2018iie}
Marchesi, S., Ajello, M., Marcotulli, L., {et~al.} 2018, \bibinfo{title}{{Compton thick AGN in the NuSTAR era},} Astrophys. J., 854, 49, \dodoi{10.3847/1538-4357/aaa410}

\bibitem[{F. Marin {et~al.}(2025)Marin, Hutsemékers, Jiang, Antonucci, Ogle, Bagul, Almeida, \& Berton}]{marin2025scatteredsynchrotronemissiongiant}
Marin, F., Hutsemékers, D., Jiang, C.~Z., {et~al.} 2025, Scattered synchrotron emission and a giant torus revealed in polarized light in the nearest radio galaxy Centaurus A, \doarXiv{2502.05002}

\bibitem[{A. {Marinucci} {et~al.}(2012){Marinucci}, {Bianchi}, {Nicastro}, {Matt}, \& {Goulding}}]{2012ApJ...748..130M}
{Marinucci}, A., {Bianchi}, S., {Nicastro}, F., {Matt}, G., \& {Goulding}, A.~D. 2012, \bibinfo{title}{{The Link between the Hidden Broad Line Region and the Accretion Rate in Seyfert 2 Galaxies},} \apj, 748, 130, \dodoi{10.1088/0004-637X/748/2/130}

\bibitem[{A. Marinucci {et~al.}(2013)Marinucci, Risaliti, Wang, Bianchi, Elvis, Matt, Nardini, \& Braito}]{Marinucci_2013}
Marinucci, A., Risaliti, G., Wang, J., {et~al.} 2013, \bibinfo{title}{X-ray absorption variability in NGC 4507,} Monthly Notices of the Royal Astronomical Society, 429, 2581–2586, \dodoi{10.1093/mnras/sts534}

\bibitem[{A. Marinucci {et~al.}(2016)Marinucci {et~al.}}]{Marinucci:2015fqo}
Marinucci, A., {et~al.} 2016, \bibinfo{title}{{NuSTAR catches the unveiling nucleus of NGC 1068},} Mon. Not. Roy. Astron. Soc., 456, L94, \dodoi{10.1093/mnrasl/slv178}

\bibitem[{M. {Markham} {et~al.}(2026){Markham}, {Bentz}, {Ferrarese}, {Onken}, \& {Vestergaard}}]{2026ApJ..1000...78M}
{Markham}, M., {Bentz}, M.~C., {Ferrarese}, L., {Onken}, C.~A., \& {Vestergaard}, M. 2026, \bibinfo{title}{{Cepheid-based Distances to NGC 4303 and NGC 1068},} \apj, 1000, 78, \dodoi{10.3847/1538-4357/ae47d8}

\bibitem[{T. {Minezaki} \& K. {Matsushita}(2015){Minezaki} \& {Matsushita}}]{2015ApJ...802...98M}
{Minezaki}, T., \& {Matsushita}, K. 2015, \bibinfo{title}{{A New Black Hole Mass Estimate for Obscured Active Galactic Nuclei},} \apj, 802, 98, \dodoi{10.1088/0004-637X/802/2/98}

\bibitem[{K. Murase(2022)Murase}]{Murase:2022dog}
Murase, K. 2022, \bibinfo{title}{{Hidden Hearts of Neutrino Active Galaxies},} Astrophys. J. Lett., 941, L17, \dodoi{10.3847/2041-8213/aca53c}

\bibitem[{K. Murase {et~al.}(2016)Murase, Guetta, \& Ahlers}]{Murase:2015xka}
Murase, K., Guetta, D., \& Ahlers, M. 2016, \bibinfo{title}{{Hidden Cosmic-Ray Accelerators as an Origin of TeV-PeV Cosmic Neutrinos},} Phys. Rev. Lett., 116, 071101, \dodoi{10.1103/PhysRevLett.116.071101}

\bibitem[{K. Murase {et~al.}(2024)Murase, Karwin, Kimura, Ajello, \& Buson}]{Murase:2023ccp}
Murase, K., Karwin, C.~M., Kimura, S.~S., Ajello, M., \& Buson, S. 2024, \bibinfo{title}{{Sub-GeV Gamma Rays from Nearby Seyfert Galaxies and Implications for Coronal Neutrino Emission},} Astrophys. J. Lett., 961, L34, \dodoi{10.3847/2041-8213/ad19c5}

\bibitem[{K. Murase {et~al.}(2020)Murase, Kimura, \& Meszaros}]{Murase:2019vdl}
Murase, K., Kimura, S.~S., \& Meszaros, P. 2020, \bibinfo{title}{{Hidden Cores of Active Galactic Nuclei as the Origin of Medium-Energy Neutrinos: Critical Tests with the MeV Gamma-Ray Connection},} Phys. Rev. Lett., 125, 011101, \dodoi{10.1103/PhysRevLett.125.011101}

\bibitem[{K. Murase {et~al.}(2026)Murase, Kimura, Mukhopadhyay, \& Bhattacharya}]{Murase:2026hrz}
Murase, K., Kimura, S.~S., Mukhopadhyay, M., \& Bhattacharya, M. 2026, \bibinfo{title}{{IceCube's convex all-sky neutrino spectrum consistent with the magnetically powered corona scenario for active galactic nuclei},} \doarXiv{2602.20145}

\bibitem[{K. Murase \& F.~W. Stecker(2023)Murase \& Stecker}]{Murase:2022feu}
Murase, K., \& Stecker, F.~W. 2023, \bibinfo{title}{{Chapter 10: High-Energy Neutrinos from Active Galactic Nuclei},} Neutrino Physics and Astrophysics, Ed. F. W. Stecker, in Encyclopedia of Cosmology II, Ed. G. G. Fazio (World Scientific), 483, \dodoi{10.1142/9789811282645_0010}

\bibitem[{E. Nardini(2017)Nardini}]{Nardini:2017qib}
Nardini, E. 2017, \bibinfo{title}{{Nuclear absorption and emission in the AGN merger NGC 6240 : the hard X-ray view},} Mon. Not. Roy. Astron. Soc., 471, 3483, \dodoi{10.1093/mnras/stx1878}

\bibitem[{A. Neronov {et~al.}(2024)Neronov, Savchenko, \& Semikoz}]{Neronov:2023aks}
Neronov, A., Savchenko, D., \& Semikoz, D.~V. 2024, \bibinfo{title}{{Neutrino Signal from a Population of Seyfert Galaxies},} Phys. Rev. Lett., 132, 101002, \dodoi{10.1103/PhysRevLett.132.101002}

\bibitem[{A. Pizzetti {et~al.}(2022)Pizzetti, Torres-Albà, Marchesi, Ajello, Silver, \& Zhao}]{Pizzetti_2022}
Pizzetti, A., Torres-Albà, N., Marchesi, S., {et~al.} 2022, \bibinfo{title}{A Multiepoch X-Ray Study of the Nearby Seyfert 2 Galaxy NGC 7479: Linking Column Density Variability to the Torus Geometry,} The Astrophysical Journal, 936, 149, \dodoi{10.3847/1538-4357/ac86c6}

\bibitem[{G. {Ponti} {et~al.}(2012){Ponti}, {Papadakis}, {Bianchi}, {Guainazzi}, {Matt}, {Uttley}, \& {Bonilla}}]{ponti_etal}
{Ponti}, G., {Papadakis}, I., {Bianchi}, S., {et~al.} 2012, \bibinfo{title}{{CAIXA: a catalogue of AGN in the XMM-Newton archive. III. Excess variance analysis},} \aap, 542, A83, \dodoi{10.1051/0004-6361/201118326}

\bibitem[{S. {Puccetti} {et~al.}(2014){Puccetti}, {Comastri}, {Fiore}, {Ar{\'e}valo}, {Risaliti}, {Bauer}, {Brandt}, {Stern}, {Harrison}, {Alexander}, {Boggs}, {Christensen}, {Craig}, {Gandhi}, {Hailey}, {Koss}, {Lansbury}, {Luo}, {Madejski}, {Matt}, {Walton}, \& {Zhang}}]{2014ApJ...793...26P}
{Puccetti}, S., {Comastri}, A., {Fiore}, F., {et~al.} 2014, \bibinfo{title}{{The Variable Hard X-Ray Emission of NGC 4945 as Observed by NuSTAR},} \apj, 793, 26, \dodoi{10.1088/0004-637X/793/1/26}

\bibitem[{C.~S. Reynolds {et~al.}(2015)Reynolds, Lohfink, Ogle, Harrison, Madsen, Fabian, Wik, Madejski, Ballantyne, Boggs, Christensen, Craig, Fuerst, Hailey, Lanz, Miller, Saez, Stern, Walton, \& Zhang}]{Reynolds_2015}
Reynolds, C.~S., Lohfink, A.~M., Ogle, P.~M., {et~al.} 2015, \bibinfo{title}{<i>NuSTAR</i>OBSERVATIONS OF THE POWERFUL RADIO-GALAXY CYGNUS A,} The Astrophysical Journal, 808, 154, \dodoi{10.1088/0004-637x/808/2/154}

\bibitem[{C. Ricci {et~al.}(2017)Ricci {et~al.}}]{Ricci:2017dhj}
Ricci, C., {et~al.} 2017, \bibinfo{title}{{BAT AGN Spectroscopic Survey - V. X-ray properties of the Swift/BAT 70-month AGN catalog},} Astrophys. J. Suppl., 233, 17, \dodoi{10.3847/1538-4365/aa96ad}

\bibitem[{J. {Rodi} {et~al.}(2026){Rodi}, {Jourdain}, {Bruni}, {Fiocchi}, {Bazzano}, {Ubertini}, {Natalucci}, \& {Roques}}]{2026A&A...710A.144R}
{Rodi}, J., {Jourdain}, E., {Bruni}, G., {et~al.} 2026, \bibinfo{title}{{NGC 1068: Exploring various geometries based on hard X-ray observations and their implications for neutrino production models in the source},} \aap, 710, A144, \dodoi{10.1051/0004-6361/202557844}

\bibitem[{R. Serafinelli {et~al.}(2024)Serafinelli, De~Rosa, Tortosa, Stella, Vagnetti, Bianchi, Ricci, Kammoun, Petrucci, Middei, Lanzuisi, Marinucci, Ursini, \& Matt}]{Serafinelli_2024}
Serafinelli, R., De~Rosa, A., Tortosa, A., {et~al.} 2024, \bibinfo{title}{Investigating the interplay between the coronal properties and the hard X-ray variability of active galactic nuclei with <i>NuSTAR</i>,} Astronomy and Astrophysics, 690, A145, \dodoi{10.1051/0004-6361/202450777}

\bibitem[{O.~I. Shuvo {et~al.}(2022)Shuvo, Johnson, Secrest, Gliozzi, Fischer, Cigan, Fernandez, \& Dorland}]{Shuvo:2022cxw}
Shuvo, O.~I., Johnson, M.~C., Secrest, N.~J., {et~al.} 2022, \bibinfo{title}{{FRAMEx. III. Radio Emission in the Immediate Vicinity of Radio-quiet AGNs},} Astrophys. J., 936, 76, \dodoi{10.3847/1538-4357/ac874a}

\bibitem[{ {Swift/BAT Catalog}(2025){Swift/BAT Catalog}}]{SwiftBAT157Catalog}
{Swift/BAT Catalog}. 2025, Swift/BAT 157-Month All-Sky Hard X-Ray Survey Catalog,, \url{https://swift.gsfc.nasa.gov/results/bs157mon/}

\bibitem[{A. Tanimoto {et~al.}(2019)Tanimoto, Ueda, Odaka, Kawaguchi, Fukazawa, \& Kawamuro}]{Tanimoto:2019tbz}
Tanimoto, A., Ueda, Y., Odaka, H., {et~al.} 2019, \bibinfo{title}{{X-Ray Spectral Model from Clumpy Torus and Its Application to Circinus Galaxy},} \dodoi{10.3847/1538-4357/ab4261}

\bibitem[{A. Tanimoto {et~al.}(2022)Tanimoto, Ueda, Odaka, Yamada, \& Ricci}]{Tanimoto:2022wrq}
Tanimoto, A., Ueda, Y., Odaka, H., Yamada, S., \& Ricci, C. 2022, \bibinfo{title}{{NuSTAR Observations of 52 Compton-thick Active Galactic Nuclei Selected by the Swift/Burst Alert Telescope All-sky Hard X-Ray Survey},} Astrophys. J. Supp., 260, 30, \dodoi{10.3847/1538-4365/ac5f59}

\bibitem[{A. Tanimoto {et~al.}(2023)Tanimoto, Wada, Kudoh, Odaka, Uematsu, \& Ogawa}]{Tanimoto:2023dzq}
Tanimoto, A., Wada, K., Kudoh, Y., {et~al.} 2023, \bibinfo{title}{{Circumnuclear Multiphase Gas in the Circinus Galaxy. V. The Origin of the X-Ray Polarization in the Circinus Galaxy},} Astrophys. J., 958, 150, \dodoi{10.3847/1538-4357/ad06ac}

\bibitem[{F. Testagrossa {et~al.}(2026)Testagrossa, Fiorillo, Comisso, Peretti, Petropoulou, \& Sironi}]{Testagrossa:2026jcs}
Testagrossa, F., Fiorillo, D. F.~G., Comisso, L., {et~al.} 2026, \bibinfo{title}{{Neutrinos and Gamma Rays from Seyfert Galaxies Constrain the Properties of Coronal Turbulence},} Astrophys. J., 1004, 4, \dodoi{10.3847/1538-4357/ae63c3}

\bibitem[{A. Tortosa {et~al.}(2017)Tortosa {et~al.}}]{Tortosa:2016loq}
Tortosa, A., {et~al.} 2017, \bibinfo{title}{{Broad-band X-ray spectral analysis of the Seyfert 1 galaxy GRS 1734{\ensuremath{-}}292},} Mon. Not. Roy. Astron. Soc., 466, 4193, \dodoi{10.1093/mnras/stw3301}

\bibitem[{J.-M. {Wang} {et~al.}(2020){Wang}, {Songsheng}, {Li}, {Du}, \& {Yu}}]{2020MNRAS.497.1020W}
{Wang}, J.-M., {Songsheng}, Y.-Y., {Li}, Y.-R., {Du}, P., \& {Yu}, Z. 2020, \bibinfo{title}{{Dynamical evidence from the sub-parsec counter-rotating disc for a close binary of supermassive black holes in NGC 1068},} \mnras, 497, 1020, \dodoi{10.1093/mnras/staa1985}

\bibitem[{J. Woo \& C.~M. Urry(2002)Woo \& Urry}]{Woo_2002}
Woo, J., \& Urry, C.~M. 2002, \bibinfo{title}{Active Galactic Nucleus Black Hole Masses and Bolometric Luminosities,} The Astrophysical Journal, 579, 530–544, \dodoi{10.1086/342878}

\bibitem[{S. Yamada {et~al.}(2024)Yamada {et~al.}}]{Yamada:2024lss}
Yamada, S., {et~al.} 2024, \bibinfo{title}{{X-Ray Winds in Nearby-to-distant Galaxies (X-WING). I. Legacy Surveys of Galaxies with Ultrafast Outflows and Warm Absorbers in z {\ensuremath{\sim}} 0{\textendash}4},} Astrophys. J. Suppl., 274, 8, \dodoi{10.3847/1538-4365/ad5961}

\bibitem[{Z.~P. Ye {et~al.}(2023)Ye {et~al.}}]{TRIDENT:2022hql}
Ye, Z.~P., {et~al.} 2023, \bibinfo{title}{{A multi-cubic-kilometre neutrino telescope in the western Pacific Ocean},} Nature Astron., 7, 1497, \dodoi{10.1038/s41550-023-02087-6}

\bibitem[{A. Zaino {et~al.}(2020)Zaino {et~al.}}]{Zaino:2020elj}
Zaino, A., {et~al.} 2020, \bibinfo{title}{{Probing the circumnuclear absorbing medium of the buried AGN in NGC 1068 through NuSTAR observations},} Mon. Not. Roy. Astron. Soc., 492, 3872, \dodoi{10.1093/mnras/staa107}

\end{thebibliography}
\bibliographystyle{aasjournalv7}
\appendix
\section{Appendix 1}
\label{appsec:details_cat}
In Table~\ref{tab:swift_bat_coordinates}, we provide additional information for the 28 X-HiNAC sources listed in Table~\ref{tab:swift_bat_sample}, including their {\em Swift}/BAT identifiers, equatorial coordinates, declinations, Eddington ratios ($\lambda_{\rm Edd}$), and optical Seyfert classifications. We also include the sample identifiers used to track the sources of the adopted X-ray measurements, as described in Section~\ref{sec:cat}.
\startlongtable
\begin{deluxetable*}{cccccccc}
\tabletypesize{\scriptsize}
\tablewidth{\textwidth}
\tablecaption{Coordinates, Eddington ratios, and optical classifications of the X-HiNAC sources. We adopt $\mathcal{R}=3$ for $\lambda_{\rm Edd} \geq 0.3$ and $\mathcal{R}=10$ for $\lambda_{\rm Edd}<0.3$.
\label{tab:swift_bat_coordinates}}
\tablehead{
\colhead{{\em Swift} ID} &
\colhead{Name} &
\colhead{R.A.} &
\colhead{Decl.} &
\colhead{$\delta\ [^\circ]$} &
\colhead{$\lambda_{\rm Edd}$} &
\colhead{Seyfert type} &
\colhead{Cat. ID}
}
\startdata
SWIFTJ0242.6+0000 & NGC 1068 & 02:42:40.70 & -00:00:47.88 & -0.0133 & 0.69823 & 2 & 123+B \\
SWIFTJ1412.9-6522 & Circinus Galaxy & 14:13:09.91 & -65:20:20.40 & -65.3390 & 0.26915 & 2 & 123+A \\
SWIFTJ1305.4-4928 & NGC 4945 & 13:05:27.48 & -49:28:05.52 & -49.4682 & 0.01377 & 2 & 1234 \\
SWIFTJ1210.5+3924 & NGC 4151 & 12:10:32.57 & +39:24:20.52 & +39.4057 & 0.03715 & 1.5--1.8\tablenotemark{a} & 12+C \\
SWIFTJ1349.3-3018 & IC 4329A & 13:49:19.27 & -30:18:33.84 & -30.3094 & 0.36392 & 1.2 & 12 (E) \\
SWIFTJ0552.2-0727 & NGC 2110 & 05:52:11.38 & -07:27:22.32 & -7.4562 & 0.13964 & 2 & 12 \\
SWIFTJ1413.2-0312 & NGC 5506 & 14:13:14.86 & -03:12:27.00 & -3.2075 & 0.21038 & 1.9/2 (NLSy1)\tablenotemark{b} & 12 (E) \\
SWIFTJ0947.6-3057 & MCG-05-23-16 & 09:47:40.15 & -30:56:56.04 & -30.9489 & 0.19231 & 1.9 & 12 (E) \\
SWIFTJ1235.6-3954 & NGC 4507 & 12:35:36.62 & -39:54:33.48 & -39.9093 & 0.14622 & 2 & 12 \\
SWIFTJ1139.0-3743 & NGC 3783 & 11:39:01.73 & -37:44:18.96 & -37.7386 & 0.19231 & 1.5 & 12 \\
SWIFTJ2201.9-3152 & NGC 7172 & 22:02:01.92 & -31:52:11.28 & -31.8698 & 0.29376 & 2 & 12+D \\
SWIFTJ0554.8+4625 & UGC 3374 & 05:54:53.62 & +46:26:21.48 & +46.4393 & 0.67608 & 1.5 & 12 \\
SWIFTJ1737.5-2908 & GRS 1734-292 & 17:37:28.34 & -29:08:02.40 & -29.1340 & 0.14060 & 1 & 12 (G) \\
SWIFTJ0048.8+3155 & Mrk 348 & 00:48:47.14 & +31:57:25.20 & +31.9570 & 0.52966 & 2 & 2 \\
SWIFTJ0615.8+7101 & Mrk 3 & 06:15:36.36 & +71:02:15.00 & +71.0375 & 0.02564 & 2 & 1234 \\
SWIFTJ1838.4-6524 & ESO 103-35 & 18:38:20.35 & -65:25:39.36 & -65.4276 & 0.36983 & 2 & 12 (E) \\
SWIFTJ1238.9-2720 & ESO 506-27 & 12:38:54.60 & -27:18:28.08 & -27.3078 & 0.07464 & 1.8 & 123 \\
SWIFTJ1023.5+1952 & NGC 3227 & 10:23:30.58 & +19:51:54.36 & +19.8651 & 0.02951 & 1.5 & 12 \\
SWIFTJ0510.7+1629 & IRAS 05078+1626 & 05:10:45.50 & +16:29:56.04 & +16.4989 & 0.53827 & 1.5 & 12 \\
SWIFTJ2044.2-1045 & Mrk 509 & 20:44:09.74 & -10:43:24.60 & -10.7235 & 1.00462 & 1.2 & 12 (E) \\
SWIFTJ1347.4-6033 & 4U1344-60 & 13:47:36.00 & -60:37:04.08 & -60.6178 & 0.31333 & 1.5\tablenotemark{c} & 12 (F) \\
SWIFTJ2304.9+1220 & NGC 7479 & 23:04:56.66 & +12:19:22.44 & +12.3229 & 0.11858 & 2 & 1234 \\
SWIFTJ2254.1-1734 & MR 2251-178 & 22:54:05.81 & -17:34:54.84 & -17.5819 & 2.28034 & 1 & 12 \\
SWIFTJ1031.7-3451 & NGC 3281 & 10:31:52.08 & -34:51:13.32 & -34.8537 & 0.01603 & 2 & 123 (E) \\
SWIFTJ1717.1-6249 & NGC 6300 & 17:16:59.47 & -62:49:14.16 & -62.8206 & 0.00543 & 2 & 12 \\
SWIFTJ2304.8-0843 & Mrk 926 & 23:04:43.46 & -08:41:08.52 & -8.6857 & 1.38357 & 1.5 & 12 \\
SWIFTJ1335.8-3416 & MCG-6-30-15 & 13:35:53.78 & -34:17:44.16 & -34.2956 & 0.04550 & 1.5 & 12 \\
SWIFTJ1442.5-1715 & NGC 5728 & 14:42:23.93 & -17:15:11.52 & -17.2532 & 0.03162 & 2 & 1234 (E) \\
\enddata
\tablecomments{
The additional samples used (also discussed in Section~\ref{subsec:details_cand}) are marked with alphabets: A -~\cite{Arevalo:2014zva}, B-~\cite{Marinucci:2015fqo}, C-~\cite{Gianolli:2023zji}, and D-~\cite{Shuvo:2022cxw}. For sources marked with E, F, and G, the intrinsic $2 - 10$ keV flux independently inferred from the {\em NuSTAR} luminosity reported by~\cite{Serafinelli_2024},~\cite{2025ApJ...994..151G}, and~\cite{Tortosa:2016loq} is shown in parentheses (see Section~\ref{sec:nustar_comparison}).
}
\tablenotetext{a}{NGC 4151 is a changing-look AGN whose optical
classification varies between approximately Seyfert 1.5 and 1.8.}
\tablenotetext{b}{NGC 5506 has historically been classified as
Seyfert 1.9/2 at optical wavelengths, while near-IR spectroscopy
reveals broad permitted lines and supports an NLSy1 classification.}
\tablenotetext{c}{4U 1344-60 has been classified as an intermediate
Seyfert, approximately Seyfert 1.5, although its precise optical
subtype is uncertain.}
\end{deluxetable*}
\clearpage
\end{document}